\documentclass{aa}
\usepackage{natbib, graphicx} 
\bibstyle{aa}     
\begin{document}
\title{Effelsberg 100-m polarimetric observations of a sample of Compact
Steep-Spectrum sources }
 
\author{ F. Mantovani  \inst{1} \and
         K.-H. Mack \inst{1} \and  
	 F.M. Montenegro-Montes \inst{1,2,3} \and
	 A. Rossetti \inst{1} \and
         A. Kraus \inst{4}}

\offprints{F. Mantovani,\\
  \email{fmantovani@ira.inaf.it}}

\institute{Istituto di Radioastronomia -- INAF, Via Gobetti 101,
 I-40129 Bologna, Italy 
\and Dpto. de Astrof\'{i}sica. Universidad de La Laguna, Avda. Astrof\'{i}sico
Fco. S\'anchez s/n, E-38200 La Laguna (Tenerife), Spain
\and Instituto de Astrof\'{i}sica de Canarias. C/ Via L\'actea s/n. E-38200
La Laguna (Tenerife), Spain
\and Max-Planck-Institut f\"ur Radioastronomie, Auf dem H\"ugel 69,
D-53121 Bonn, Germany
}

\date{Received \today; accepted ???}

\abstract
% Context heading
{}
%Aims heading
{We completed observations with the Effelsberg 100-m radio telescope 
to measure the polarised 
emission from a complete sample of Compact Steep-Spectrum sources and
improve our understanding of the physical conditions inside and around 
regions of radio emission embedded in dense interstellar environments.}
%Methods heading
{We observed the sources at
four different frequencies, namely 2.64\,GHz, 4.85\,GHz, 8.35\,GHz, and
10.45\,GHz, making use of the polarimeters available at the Effelsberg 
telescope. We complemented these measurements with polarisation parameters at 
1.4~GHz derived from the NRAO VLA Sky Survey. Previous  single dish 
measurements were taken from the catalogue of Tabara and Inoue.}
%Results heding
{The depolarisation index 
DP was computed for four pairs of frequencies. A drop in the fractional 
polarisation appeared in the radio emission when observing at frequencies below 
$\sim$2\,GHz. Rotation measures were derived for about 25\% of the sources 
in the sample. The values, in the source rest frame, range from about 
--20 rad m$^{-2}$ found for 3C138 to 3900  rad m$^{-2}$ in 3C119.
In all cases, the $\lambda^2$ law is closely followed.}
% Conclusions heading 
{
The presence of a foreground screen as predicted by the Tribble model
or with ``partial coverage'' as defined by ourselves 
can explain the polarimetric behaviour of the CSS sources detected in 
polarisation by the present observations. Indication of repolarisation
at lower frequencies was found for some sources. A case of
possible variability in the fractional polarisation is also suggested.
The most unexpected result was found for the distribution 
of the fractional polarisations versus the linear sizes of the sources. 
Our results appear to disagree with the findings
of Cotton and collaborators and Fanti and collaborators for the B3-VLA sample 
of CSS sources, the so-called ``Cotton effect'', i.e., a strong drop in 
polarised intensity for the most compact sources below a given frequency. This
apparent contradiction may, however, be caused by the large contamination 
of the sample by quasars with respect to the B3-VLA. }

\titlerunning{Effelsberg polarimetry of CSS sources}
\maketitle 
\section{Introduction} 

Measurements of the polarised emission from Giga-Hertz-Peaked Spectrum (GPS)
and Compact Steep-Spectrum (CSS) sources can provide important information
about the physical conditions inside and around the region of radio emission.
GPS/CSS sources are physically small objects with radio sizes smaller
than 1 kpc (essentially on the Narrow Line Region size scale) and 15 kpc,
respectively,  that reside inside their host galaxies. The most widely 
accepted
interpretation  is that GPS/CSS galaxies are young radio sources
(of ages $< 10^6$\,yr, Fanti et al. 1990).
This view is supported by measurements of hot spot advance speeds (e.g., 
Polatidis \& Conway 2003) and by means of spectral ageing studies (Murgia et 
al. 1999). GPS sources, which are the youngest, will grow 
into CSS and eventually into classical extended radio sources.
Consequently, the measurement of the physical properties of GPS and CSS 
sources can provide insight into the conditions at  
the birth of a powerful radio source and those of sources
developing in dense interstellar environments (see Fanti et al. 1995 and 
Readhead et al. 1996). The observations suggest that intrinsic distorsions
in CSS sources are be caused by their interactions with dense and inhomogeneous
gaseous environments. Asymmetries in terms of flux density,
arm ratio, spectral index, and polarisation suggest that they are expanding
through the dense inhomogeneous interstellar medium of their host
galaxies (e.g., Saikia et al. 2003; Rossetti et al. 2006). 

An effect on the synchrotron polarised emission produced by this magnetized 
thermal  plasma is
Faraday rotation, which is proportional to the product of electron density
and the magnetic field component parallel to the direction of propagation
integrated along the line of sight. The Rotation Measure (RM) is the
amount of  Faraday rotation expressed in rad m$^{-2}$.
The Faraday rotation can unambiguously be determined from observations for 
at least three
different wavelengths. Strong variations in Faraday rotation across the
telescope beam will reduce or depolarise the observed fractional polarisation.
This effect can often completely depolarise regions of emission.

In many sources, the polarisation increases with increasing frequency
when observed at similar angular resolutions. The fractional
polarisations of CSS sources at 15 GHz (6--7 \%) indeed tend to be higher
than at 5 GHz (1--3 \%; van Breugel et al. 1984; Saikia et al. 1987).
This suggests that large Faraday depths are responsible
for the depolarisation between 15 GHz and 5 GHz rather than magnetic field
geometry.  

The magnetized plasma responsible for Faraday rotation and 
depolarisation can be situated either within the source, in a foreground screen,
or in both. However, there are several indications, such as the magnitudes of 
the RMs, the total rotation of the electric vector position angle $>90^\circ$ 
without very high depolarisation, and the lack of correlation between high RMs 
in extragalactic sources and Galactic latitude (see for example the 
observations of Cygnus A by Dreher et al. 1987), that most of the observable 
effects are not internal to the source, but are produced in foreground material
in the vicinity of the radio synchrotron source (e.g., Leahy 1990).

A model to describe the depolarisation behaviour in external screens has been
discussed by Burn (1966) and later generalized by Tribble (1991) who considered
the importance of the sizes of the individual Faraday cells in relation to the
observing beam sizes. A variation in the ratio of these sizes can cause a 
more or less strong decline in the intrinsic polarisation percentage
towards longer wavelengths. A successful application of the Tribble model, 
which implies the existence of a foreground screen, can be found in Fanti et 
al. (2004), who analysed multi-frequency observations of the B3-VLA CSS sources 
(Fanti et al. 2001).

Many GPS/CSS sources have been observed mainly with radio
interferometers. These
investigations have provided very interesting results about the polarised
state of the radio emission in GPS/CSS sources. A summary can be found in
O'Dea (1998) and Cotton et al. (2003).
However, apart from the B3-VLA sample (Fanti et al. 2001), most of the samples 
observed in polarisation so far (L\"udke et al.
1998; Akujor et al. 1995; van Breugel et al. 1984; Rossetti et al. 2008) 
contained subsets of sources whose incompleteness hamper statistical studies 
or which were observed at a single or dual frequency.

The set up of this paper is as follows. Sample selection, observations and 
data reduction are described in Sect.\,2. Results from the observations are 
presented in Sect.\,3 and discussed in Sect\,4. In Sect.\,5 we draw our 
conclusions. Plots of the fractional polarisation $m$ and Rotation Measures RM 
versus wavelength squared are compiled in Appendix 1. Comments about individual 
sources can be found in Appendix 2.
\section{Observations and data reduction} 
\label{sec:observation}

\subsection{The sample}

The sample of CSS sources subject to the present investigation was  
constructed 
by Fanti et al. (1990) from the 3CR catalogue (Jenkins et al. 1977) and 
the Peacock \& Wall (1982, hereafter PW) sample. Since
there is enough information about the radio structures and the spectral
behaviour of the sources in the 3CR and PW catalogues, it is safe to assume
that the sample contains all the CSS sources belonging to those catalogues
with projected linear sizes less than 15 kpc, flux density at 178 MHz
$\ge$ 10 Jy, log P$_{178} > $26.5 W Hz$^{-1}$, and with 
$|b|> 10^\circ, \delta > 10^\circ$, and can thus be considered statistically 
complete.

We observed all the sources belonging to this sample
with the Effelsberg 100-m radio telescope at four frequencies between
2.64\,GHz and 10.45\,GHz in a relatively short period of time
(i.e., contemporaneously at the three higher frequencies, and 16 months later 
at 2.64\,GHz). We choose this approach to avoid time 
variability effects when measuring the percentage of polarised flux
density for each of these sources, the depolarisation indices, and the RMs 
with the aim of understanding more clearly the physical conditions in which CSS
sources eventually expand to become larger sources.  

Polarisation parameters at 1.4~GHz were also derived for the full sample from
the National Radio Astronomy Observatory
Very Large Array Sky Survey (NVSS; Condon et al. 1998). Because of their small
angular sizes, CSS sources are point-like at the NVSS resolution.
In the analysis of sources that were found polarised at two or more of the five 
frequencies above, we complemented the polarisation measurements with 
those listed by Tabara \& Inoue (1980).

For consistency with previous work, we used the
cosmology H$_0=100$ km\,s$^{-1}$ Mpc$^{-1}$ and q$_0=0.5$.   

\subsection{Observations}

The Effelsberg 100-m telescope was used to observe the complete
sample of GPS/CSS sources constructed by Fanti et al. (1990). 
To minimize the occurrence of possible n$\pi$ ambiguities in the
determination of RMs, we observed the sources at
four independent frequencies, namely 2.64\,GHz, 4.85\,GHz, 8.35\,GHz, and
10.45\,GHz, making use of the polarimeters available at the Effelsberg 
telescope. The observations were carried
out in the period January 26 to February 1, 2005 at 4.85\,GHz, 8.35\,GHz, 
and 10.45\,GHz and June 24 to 26, 2006 at 2.64\,GHz.
Since all the target sources are point-like to the Effelsberg telescope beams,
we used cross-scanning to determine the total intensity and polarisation 
characteristics. All sources in the sample are very bright, thus standard
cross-scans along the azimuth and elevation axes were used, with 
4 to 8 subscans, according to the source flux densities. Table\,1 summarises 
the observing parameters. For further
details about the observation mode and a description of the receivers, we
refer to Montenegro--Montes et al. (2008), Klein et al. (2003), and
references therein.
\tabcolsep0.1cm
\begin{table}[h]
\centering
\caption{ Observing parameters and estimated integration times}
\begin{tabular}{lrrrr}
\hline
Centre Frequency [GHz] & 2.64 &  4.85         &  8.35    &  10.45   \\
\hline
Bandwidth [MHz]        & 80   &    500        &  1100    &  300    \\
System Temp.(zenith) [K] & 17 & 27            & 22       & 53        \\
Scan length [$'$]      & 16   &  12           &  8       &  6        \\
Scan speed [$'$/min]   & 45   &   45          & 40       &  30      \\
% Scan duration [s]      & 30   &     30        &    26    &  26         \\
\hline
\end{tabular}
\end{table}
The calibration sources 3C\,286 and 3C\,295 were regularly observed to correct 
for time-dependent gain instabilities and to bring our measurements onto an 
absolute flux density scale (Baars et al. 1977). The quasar 3C\,286 was also 
used as a polarisation calibrator to obtain
the polarisation degree $m$ and the polarisation angle $\chi$ 
in agreement with values in the literature (Tabara et al. 1980). 
The unpolarised 
planetary nebula NGC\,7027 was also observed to estimate the
instrumental polarisation.  We found an instrumental polarisation
($p_{instr}$)
of 0.5\% at 2.64\,GHz, 0.5\% at 4.85\,GHz, 0.3\% at 8.35\,GHz, and 0.8\% at 
10.45\,GHz. 

The measurement of flux densities from the single-dish cross-scans was done by 
fitting Gaussians to the signal of
the polarimeter output channels (Stokes I, Q and U) and identifying the 
Gaussian amplitudes with the flux 
densities $S_{I}$, $S_{Q}$, and $S_{U}$. For all sources with significant 
$S_{Q}$ and $S_{U}$ contributions, the 
polarised flux density $S_{P}$, the degree of linear polarisation $m$, and the 
polarisation angle $\chi$ were computed. 

We consider three main contributions to the flux density error as in 
Klein et al. (2003). These are 

(i) the calibration 
error $\Delta S_{c}$, which is estimated to be the dispersion in the different 
observations of the flux density 
calibrators, i.e., about 2\% at all observing frequencies; 

(ii) the error introduced by noise, $\Delta S_{i}\sim$2\,mJy
({\it i} = I,Q,U), which is estimated from the noise at the scan edges;  

(iii) the confusion error $\Delta S_{conf}$ caused by background 
sources within the beam area, estimated to be 1.5\,mJy at 2.64\,GHz, 
0.45\,mJy at 4.85\,GHz, 0.17\,mJy at 8.35\,GHz, and 0.08\,mJy at 10.45\,GHz
(Klein et al. 2003; the value at 8.35\,GHz was extrapolated from
these existing measurements. The confusion limits can be neglected in the
calculation of the total error in Stokes $Q$ and $U$.

These contribute to the total error in the following way:

$$
\Delta S_{\it i} = \sqrt{(S_{\it i} * \Delta S_{c})^2 + \Delta S_{\it i}^2 + \Delta S_{conf}^2} $$

\noindent
Since we are dealing with relatively bright targets, instrumental polarisation 
is an issue in many cases. This has been included in the error of the 
fractional polarisation in the following way:

$$
\Delta S_P = \sqrt{\frac{(S_Q * \Delta S_Q)^2 + (S_U * \Delta S_U)^2}{S_P^2}
+ (p_{instr} * S_I)^2}
$$

\noindent
For all other errors we follow the definitions given by Klein et al. (2003). 
The errors associated with the position angles $\chi$ also account for the 
distribution, assumed to be Gaussian, in the values of $\chi$ obtained for 
the calibrator 3C286 in the calibration process.
\subsection{Archival data}
\label{subsubsec:literature}

The Effelsberg measurements were complemented
with data of the NRAO VLA Sky Survey (NVSS) at 1.4\,GHz 
(Condon et al. 1998). In all these measurements, our targets are 
point-like to the corresponding beams, thus beam depolarising effects 
are avoided. The polarised flux densities in Table 2 are given for sources with
polarised flux densities greater than three times the rms error. The rms 
uncertainty is computed following Eq. 49 in Condon et al. (1998)

$$
\sigma_P^2 \approx 2\sigma_{Q,U}^2 + \epsilon_P^2 A_P^2
$$ 

\noindent
where $\epsilon_P^2$ is the residual instrumental polarisation, which is
about 0.12\% for a large sample of sources stronger than 1\,Jy and $A_P$
is the fitted peaked amplitude. In Table 3, we report the fractional 
polarisation values for these sources.

%
%\newpage
%
\tabcolsep0.05mm
\begin{table*}[h]
\centering
\caption{Flux densities and polarised flux densities from Effelsberg 100-m and 
NVSS measurements.}
\tiny
\begin{tabular}{llrcrrcrrcrrcrrcrrcrrcrrcrrcrrcr}
\hline
\multicolumn{2}{c}{Name} & \multicolumn{3}{c}{S$_{1.4}$} & \multicolumn{3}{c}{S$_{2.64}$} & \multicolumn{3}{c}{S$_{4.85}$} & \multicolumn{3}{c}{S$_{8.35}$} & \multicolumn{3}{c}{S$_{10.45}$} & \multicolumn{3}{c}{S$_{p 1.4}$} & \multicolumn{3}{c}{S$_{p 2.64}$} &\multicolumn{3}{c}{S$_{p 4.85}$} & \multicolumn{3}{c}{S$_{p 8.35}$} & \multicolumn{3}{c}{S$_{p 10.45}$ }  \\
&  & \multicolumn{3}{c}{[mJy]}& \multicolumn{3}{c}{[mJy]}& \multicolumn{3}{c}{[mJy]} & \multicolumn{3}{c}{[mJy]}  &  \multicolumn{3}{c}{[mJy]}       &         \multicolumn{3}{c}{[mJy]}  & \multicolumn{3}{c}{[mJy]} &\multicolumn{3}{c}{[mJy]}& \multicolumn{3}{c}{[mJy]} & \multicolumn{3}{c}{[mJy]} \\
\hline
3C43       &~0127+23 &~ 29414 &$\pm$&  88 &~  1787&$\pm$& 37 &~ 1109&$\pm$& 22 &~  698&$\pm$& 14 &~   578 &$\pm$&12     &~~~~~    & $<$10.8 &                   &~~~~~      & $<$ 38.4 &       &~~~~~      & $<$6.6     &      & \multicolumn{3}{c}{~ 14.7 $\pm$  4.1 } & \multicolumn{3}{c}{~12.2$\pm$ 2.3 }  \\
3C48       &~0134+32 &~ 16018 &$\pm$& 481 &~  9373&$\pm$&191 &~ 5534&$\pm$&113 &~ 3243&$\pm$& 65 &~  2495 &$\pm$&52     &\multicolumn{3}{c}{~70.8 $\pm$ 57.6} &~      & $<$195.0 &       & \multicolumn{3}{c}{~185.5  $\pm$ 43.1 }&\multicolumn{3}{c}{~ 160.5 $\pm$  18.8}  &\multicolumn{3}{c}{~139.5 $\pm$5.6 }  \\
3C49       &~0138+13 &~  2740 &$\pm$&  82 &~  1552&$\pm$& 33 &~  878&$\pm$& 18 &~  494&$\pm$& 10 &~   376 &$\pm$&8      &      & $<$9.9  &                   &~      & $<$ 34.5 &       &~       & $<$20.4    &      &~~~~~      &$<$8.7    &       &~~~~~      & $<$ 6.3    &       \\
3C67       &~0221+27 &~  3024 &$\pm$&  91 &~  1757&$\pm$& 36 &~  997&$\pm$& 20 &~  575&$\pm$& 12 &~   419 &$\pm$&9      &\multicolumn{3}{c}{~17.0 $\pm$ 10.9} &~      & $<$ 37.8 &       &~       & $<$23.4    &      &\multicolumn{3}{c}{~ 22.3 $\pm$  3.4 } & \multicolumn{3}{c}{~15.7$\pm$ 1.9 }  \\
4C34.07    &~0223+34 &~  2894 &$\pm$&  87 &~  2483&$\pm$& 51 &~ 2234&$\pm$& 46 &~ 1452&$\pm$& 30 &~  1212 &$\pm$&25     &~     & $<$10.5 &                   &~      & $<$ 51.6 &       &~       & $<$52.2    &      &~       &$<$24.9   &       &~      & $<$13.2    &       \\
4C16.09    &~0316+16 &~  8028 &$\pm$& 241 &~  5038&$\pm$&103 &~ 2937&$\pm$& 60 &~ 1563&$\pm$& 32 &~  1187 &$\pm$&25     &~     & $<$28.9 &                   &~      & $<$103.5 &       &~       & $<$6.9     &      &~       &$<$27.0   &       &~      & $<$12.9    &       \\
           &~0319+12 &~  1907 &$\pm$&  67 &~  1704&$\pm$& 35 &~ 1791&$\pm$& 37 &~ 1527&$\pm$& 31 &~  1549 &$\pm$&32     &\multicolumn{3}{c}{~ 71.8$\pm$ 7.0}  & \multicolumn{3}{c}{~95.3 $\pm$  12.3}  &\multicolumn{3}{c}{~113.0$\pm$  14.1} & \multicolumn{3}{c}{~88.9 $\pm$  8.9}   & \multicolumn{3}{c}{~77.2$\pm$ 5.7}   \\
3C93.1     &~0345+33 &~  2365 &$\pm$&  71 &~  1390&$\pm$& 29 &~  926&$\pm$& 19 &~  477&$\pm$& 10 &~   383 &$\pm$&8      &~     & $<$8.6  &                   &~      & $<$ 30.0 &       &~       & $<$21.6    &      &~       &$<$8.4    &       &  \multicolumn{3}{c}{~9.1$\pm$ 1.6}   \\ 
4C76.03    &~0404+76 &~  5619 &$\pm$& 169 &~  4042&$\pm$& 83 &~ 3355&$\pm$& 69 &~ 2087&$\pm$& 42 &~  1648 &$\pm$&34     &~     & $<$20.3 &                   &~      & $<$ 83.4 &       &~       & $<$78.3    &      &~       &$<$36.0   &       &~      & $<$18.0    &       \\ 
           &~0428+20 &~  3755 &$\pm$& 112 &~  3166&$\pm$& 65 &~ 2761&$\pm$& 57 &~ 1542&$\pm$& 31 &~  1289 &$\pm$&26     &~     & $<$13.6 &                   &~      & $<$ 65.7 &       &~       & $<$64.5    &      &~       &$<$26.7   &       &~      & $<$14.1    &       \\
3C119      &~0429+41 &~  9832 &$\pm$& 295 &~  6126&$\pm$&126 &~ 4722&$\pm$& 97 &~ 2677&$\pm$& 54 &~  2260 &$\pm$&47     &~     & $<$35.4 &                   &~      & $<$125.4 &       &~       & $<$110.1   &      & \multicolumn{3}{c}{~157.9 $\pm$ 15.7}  &\multicolumn{3}{c}{~200.6$\pm$  8.9 }  \\
3C138      &~0518+165&~  8603 &$\pm$& 258 &~  5780&$\pm$&119 &~ 4119&$\pm$& 84 &~ 2552&$\pm$& 53 &~  2160 &$\pm$&45     &\multicolumn{3}{c}{~619.0 $\pm$ 31.0}& \multicolumn{3}{c}{~595.5$\pm$  41.0}  &\multicolumn{3}{c}{~ 411.6 $\pm$ 32.8}   &\multicolumn{3}{c}{~297.8 $\pm$ 17.6}  & \multicolumn{3}{c}{~218.8$\pm$8.8 }  \\
3C147      &~0538+49 &~ 22880 &$\pm$& 686 &~ 13436&$\pm$&276 &~10120&$\pm$&207 &~ 4735&$\pm$& 96 &~  3823 &$\pm$&80     &~     & $<$82.3 &                   &~      & $<$274.5 &       &~       & $<$236.1   &      &~       &$<$81.6   &       &\multicolumn{3}{c}{~47.3 $\pm$ 13.6 }  \\
3C186      &~0740+38 &~  1236 &$\pm$&  37 &~   615&$\pm$& 14 &~  271&$\pm$&  8 &~  114&$\pm$&  3 &~    85 &$\pm$&3      &~     & $<$4.6  &                   &~      & $<$ 15.3 &       &~       & $<$6.3     &      &~       &$<$2.4    &       &~      & $<$ 3.0    &       \\ 
3C190      &~0758+14 &~  2734 &$\pm$&  82 &~  1359&$\pm$& 28 &~  746&$\pm$& 16 &~  426&$\pm$& 10 &~   360 &$\pm$&8      &~     & $<$9.9  &                   &~      & $<$ 31.5 &       &~       & $<$17.4    &      &~       &$<$7.8    &       &~      & $<$ 5.1    &       \\ 
3C216      &~0906+43 &~  4239 &$\pm$& 127 &~  1446&$\pm$& 50 &~ 1731&$\pm$& 35 &~ 1413&$\pm$& 29 &~  1337 &$\pm$&29     &~     & $<$15.3 &                   &~      & $<$ 52.5 &       &~       & $<$40.5    &      &~       &$<$24.3   &       &~      & $<$14.7    &       \\ 
3C237      &~1005+07 &~  6522 &$\pm$& 196 &~  3608&$\pm$& 74 &~ 1971&$\pm$& 42 &~ 1034&$\pm$& 21 &~   823 &$\pm$&17     &~     & $<$23.5 &                   &~      & $<$ 76.5 &       &~       & $<$45.9    &      &~       &$<$18.0   &       & \multicolumn{3}{c}{~ 9.7$\pm$ 3.2 }  \\ 
3C241      &~1019+22 &~  1686 &$\pm$&  51 &~   799&$\pm$& 17 &~  350&$\pm$&  7 &~  161&$\pm$& 33 &~   112 &$\pm$&3      &~     & $<$6.2  &                   &~      & $<$ 19.5 &       &~       & $<$8.1     &      &~       &$<$3.0    &       &~      & $<$ 1.2    &       \\ 
4C31.38    &~1153+31 &~  2978 &$\pm$&  89 &~  1751&$\pm$& 36 &~ 1030&$\pm$& 21 &~  545&$\pm$& 11 &~   441 &$\pm$&10     &~     & $<$10.8 &                   &~      & $<$ 45.0 &       &~       & $<$24.0    &      &~       &$<$9.6    &       &  \multicolumn{3}{c}{~ 8.6$\pm$  1.8 }  \\ 
3C268.3    &~1203+64 &~  3719 &$\pm$& 112 &~  1947&$\pm$& 39 &~ 1153&$\pm$& 23 &~  618&$\pm$& 13 &~   479 &$\pm$&10     &~     & $<$13.4 &                   &~      & $<$ 40.8 &       &~       & $<$27.0    &      & \multicolumn{3}{c}{~17.2 $\pm$  3.7 } & \multicolumn{3}{c}{~ 14.2$\pm$   2.1 }  \\ 
           &~1225+36 &~  2098 &$\pm$&  63 &~  1434&$\pm$& 30 &~  753&$\pm$& 16 &~  353&$\pm$&  9 &~   239 &$\pm$&5      &~     & $<$7.7  &                   &~      & $<$ 31.5 &       &~       & $<$17.7    &      &~       &$<$6.3    &       &~      & $<$2.4     &       \\ 
3C277.1    &~1250+56 &~  2288 &$\pm$&  69 &~  1343&$\pm$& 30 &~  790&$\pm$& 16 &~  481&$\pm$& 10 &~   384 &$\pm$&8      &~     & $<$8.3  &                   &~      & $<$ 29.4 &       &~       & $<$18.3    &      & \multicolumn{3}{c}{~14.3 $\pm$  2.9 } & \multicolumn{3}{c}{~  9.9$\pm$   1.8 }  \\ 
4C32.44    &~1323+32 &~  4862 &$\pm$& 146 &~  3369&$\pm$& 70 &~ 2293&$\pm$& 47 &~ 1540&$\pm$& 31 &~  1293 &$\pm$&28     &~     & $<$17.5 &                   & \multicolumn{3}{c}{~17.3 $\pm$   2.3}  &~       & $<$53.4    &      &~       &$<$26.7   &       &~      & $<$14.1    &       \\ 
3C286      &~1328+30 &~ 14902 &$\pm$& 447 &~ 10607&$\pm$&219 &~ 7430&$\pm$&152 &~ 5179&$\pm$&104 &~  4474 &$\pm$&94     &\multicolumn{3}{c}{~999.9 $\pm$ 53.7}&\multicolumn{3}{c}{~1152.7$\pm$  74.1}  & \multicolumn{3}{c}{~776.7 $\pm$  59.4} &\multicolumn{3}{c}{~607.4 $\pm$ 31.5}  &\multicolumn{3}{c}{~525.8$\pm$ 18.4 }  \\ 
3C287      &~1328+27 &~  7052 &$\pm$& 212 &~  4697&$\pm$& 97 &~ 3130&$\pm$& 64 &~ 2050&$\pm$& 42 &~  1748 &$\pm$&36     &\multicolumn{3}{c}{~ 45.6 $\pm$ 25.4}&\multicolumn{3}{c}{~168.7$\pm$  32.3}  &\multicolumn{3}{c}{~ 116.6 $\pm$ 24.4} &\multicolumn{3}{c}{~ 65.2 $\pm$ 11.8 } & \multicolumn{3}{c}{~ 58.6$\pm$  6.4 }  \\
4C62.22    &~1358+62 &~  4308 &$\pm$& 129 &~  2765&$\pm$& 57 &~ 1701&$\pm$& 35 &~ 1080&$\pm$& 22 &~   888 &$\pm$&18     &~     & $<$15.6 &                   &~      & $<$ 57.3 &       &~       & $<$39.6    &      &~       &$<$18.6   &       &~      & $<$9.9     &       \\
           &~1413+34 &~  1864 &$\pm$&  56 &~  1488&$\pm$& 30 &~ 1059&$\pm$& 22 &~  735&$\pm$& 15 &~   621 &$\pm$&14     &~     & $<$6.8  &                   & \multicolumn{3}{c}{~45.8 $\pm$   11.2}  &~       & $<$24.6    &      &~       &$<$12.9   &       &~      & $<$7.5     &       \\   
3C298      &~1416+06 &~  6100 &$\pm$& 183 &~  2888&$\pm$& 60 &~ 1432&$\pm$& 30 &~  789&$\pm$& 17 &~   616 &$\pm$&14     &~     & $<$22.0 &                   &~      & $<$ 60.0 &       &~       & $<$33.3    &      & \multicolumn{3}{c}{~14.6 $\pm$ 4.6} & \multicolumn{3}{c}{~12.8 $\pm$ 2.6 }  \\ 
3C299      &~1419+41 &~  3147 &$\pm$& 111 &~  1721&$\pm$& 35 &~  922&$\pm$& 20 &~  502&$\pm$& 10 &~   400 &$\pm$&9      &~     & $<$11.4 &                   &~      & $<$ 40.8 &       &~       & $<$21.6    &      &~       &$<$9.0    &       &~      & $<$6.6     &       \\   
OQ172      &~1442+10 &~  2418 &$\pm$&  72 &~  1398&$\pm$& 38 &~ 1032&$\pm$& 21 &~  641&$\pm$& 14 &~   516 &$\pm$&11     &\multicolumn{3}{c}{~31.7$\pm$   8.8} & \multicolumn{3}{c}{~63.2 $\pm$  12.0}  & \multicolumn{3}{c}{~26.5  $\pm$  8.0} & \multicolumn{3}{c}{~ 14.2 $\pm$ 3.8 }   &~      & $<$8.7     &       \\
3C303.1    &~1443+77 &~  1880 &$\pm$&  66 &~   882&$\pm$& 19 &~  413&$\pm$&  8 &~  199&$\pm$&  4 &~   128 &$\pm$&3      &~     & $<$6.9  &                   &~      & $<$ 20.1 &       &~       & $<$9.6     &      &~       &$<$3.9    &       &~      & $<$2.4     &       \\ 
3C305.1    &~1447+77 &~  1666 &$\pm$&  50 &~   837&$\pm$& 18 &~  412&$\pm$&  9 &~  208&$\pm$&  4 &~   154 &$\pm$&3      &~     & $<$6.2  &                   &~      & $<$ 20.1 &       &~       & $<$9.6     &      &~       &$<$11.7   &       &~      & $<$3.9     &       \\
3C309.1    &~1458+71 &~  7468 &$\pm$& 224 &~  5129&$\pm$&105 &~ 3552&$\pm$& 73 &~ 2568&$\pm$& 52 &~  2196 &$\pm$&46     &\multicolumn{3}{c}{~91.1 $\pm$  26.9}& \multicolumn{3}{c}{~188.3 $\pm$  35.2}  &  \multicolumn{3}{c}{~8.5  $\pm$   2.8} &\multicolumn{3}{c}{~70.8 $\pm$14.8}  &\multicolumn{3}{c}{~55.5 $\pm$ 8.8 }  \\
3C318      &~1517+20 &~  2688 &$\pm$&  81 &~  1417&$\pm$& 29 &~  764&$\pm$& 16 &~  417&$\pm$&  8 &~   322 &$\pm$&7      &~     & $<$9.8  &                   &~      & $<$ 30.9 &       & \multicolumn{3}{c}{~25.8  $\pm$   6.0} &\multicolumn{3}{c}{~ 27.6 $\pm$ 2.6}  &\multicolumn{3}{c}{~29.6 $\pm$ 1.8 }  \\
           &~1600+33 &~  2991 &$\pm$&  90 &~  1980&$\pm$& 41 &~ 1433&$\pm$& 29 &~ 1049&$\pm$& 21 &~   915 &$\pm$&20     &~     & $<$10.8 &                   &~      & $<$ 42.3 &       &~       & $<$33.3    &      &~       &$<$18.3   &       &~      & $<$10.5    &       \\
OS111      &~1607+26 &~  4908 &$\pm$& 147 &~  3140&$\pm$& 65 &~ 1728&$\pm$& 35 &~  902&$\pm$& 18 &~   681 &$\pm$&15     &~     & $<$17.7 &                   &~      & $<$ 98.7 &       &~       & $<$40.2    &      &~       &$<$15.6   &       &~      & $<$8.1     &       \\   
3C343      &~1634+62 &~  5002 &$\pm$& 150 &~  2858&$\pm$& 59 &~ 1503&$\pm$& 31 &~  808&$\pm$& 16 &~   622 &$\pm$&13     &~     & $<$18.0 &                   &~      & $<$ 59.1 &       &~       & $<$35.1    &      &~       &$<$14.1   &       & \multicolumn{3}{c}{~8.4 $\pm$ 2.5 }  \\   
3C343.1    &~1637+62 &~  4611 &$\pm$& 138 &~  2384&$\pm$& 49 &~ 1190&$\pm$& 24 &~  631&$\pm$& 13 &~   479 &$\pm$&10     &~     & $<$16.6 &                   &~      & $<$ 49.2 &       &~       & $<$27.9    &      &~       &$<$11.1   &       &~      &  $<$6.0    &       \\
3C346      &~1641+17 &~  3666 &$\pm$& 110 &~  2260&$\pm$& 47 &~ 1426&$\pm$& 29 &~  915&$\pm$& 19 &~   773 &$\pm$&17     &\multicolumn{3}{c}{~80.6 $\pm$  13.3}&~      & $<$ 47.7 &       &~       & $<$33.3    &      &~       &$<$45.9   &       &\multicolumn{3}{c}{~ 13.8 $\pm$ 3.0 }  \\  
4C39.56    &~1819+39 &~  3507 &$\pm$& 105 &~  1871&$\pm$& 42 &~  938&$\pm$& 19 &~  463&$\pm$& 10 &~   337 &$\pm$&7      &~     & $<$12.7 &                   &~      & $<$ 39.9 &       &~       & $<$21.9    &      &~       &$<$8.1    &       &~      &  $<$4.5    &       \\
3C380      &~1828+48 &~ 13573 &$\pm$& 413 &~  8139&$\pm$&167 &~ 5073&$\pm$&105 &~ 3523&$\pm$& 71 &~  3055 &$\pm$&63     &\multicolumn{3}{c}{~62.2$\pm$   49.5}&~      &$<$ 166.5 &       &~       & $<$118.2   &      &~       &$<$60.6   &       &~      &  $<$32.7   &       \\  
4C29.56    &~1829+29 &~  2924 &$\pm$&  88 &~  1945&$\pm$& 40 &~ 1168&$\pm$& 24 &~  679&$\pm$& 14 &~   533 &$\pm$&12     &~     & $<$10.6 &                   &~      & $<$ 40.8 &       &~       & $<$27.3    &      &~       &$<$12.0   &       &~      &  $<$6.9    &       \\
4C11.69    &~2230+11 &~  7202 &$\pm$& 216 &~  5695&$\pm$&118 &~ 4135&$\pm$& 85 &~ 3396&$\pm$& 69 &~  3209 &$\pm$&67     &\multicolumn{3}{c}{~114.0 $\pm$26.0} & \multicolumn{3}{c}{~229.6$\pm$  39.1}  & \multicolumn{3}{c}{~147.6 $\pm$ 32.3} &\multicolumn{3}{c}{~ 127.3 $\pm$ 19.6 } &\multicolumn{3}{c}{~ 116.4 $\pm$11.6 }  \\ 
3C454      &~2249+18 &~  2133 &$\pm$&  64 &~  1265&$\pm$& 26 &~  761&$\pm$& 16 &~  469&$\pm$& 10 &~   383 &$\pm$&8      &\multicolumn{3}{c}{~77.4 $\pm$  7.8} & \multicolumn{3}{c}{~73.6$\pm$     9.1}  & \multicolumn{3}{c}{~ 71.0 $\pm$ 6.1} &\multicolumn{3}{c}{~ 52.7 $\pm$  2.9 } &\multicolumn{3}{c}{~ 42.8 $\pm$  1.9 }  \\ 
3C454.1    &~2248+71 &~  1555 &$\pm$&  47 &~   696&$\pm$& 15 &~  287&$\pm$&  6 &~  119&$\pm$&  3 &~    87 &$\pm$&2      &~     & $<$5.8  &                   &~      & $<$ 16.8 &       &~       & $<$6.9     &      &~       &$<$3.6    &       &~      & $<$0.9     &       \\ 
3C455      &~2252+12 &~  2706 &$\pm$&  81 &~  1491&$\pm$& 30 &~  772&$\pm$& 16 &~  421&$\pm$& 93 &~   312 &$\pm$&7      &\multicolumn{3}{c}{~55.8 $\pm$  9.9} &\multicolumn{3}{c}{~50.5$\pm$    10.9}  &~       & $<$18.0    &      &~       &$<$7.8    &       & \multicolumn{3}{c}{~4.7$\pm$ 1.5 }  \\   
           &~2342+82 &~  3777 &$\pm$& 113 &~  2275&$\pm$& 47 &~ 1308&$\pm$& 27 &~  747&$\pm$& 15 &~   579 &$\pm$&12     &~     & $<$13.7 &                   &~      & $<$ 47.4 &       &~       & $<$30.6    &      &~       &$<$13.2   &       &~      & $<$6.6     &       \\  
\hline
%
%\normalfont
%
\end{tabular}
\normalfont
\smallskip\noindent
\flushleft{\normalsize {
The values at the various observing frequencies (in GHz) 
are organised as follows: column 1, source name; column 2, other name; 
columns 3 to 7, flux density S; columns 8 to 12, polarised flux density S$_p$.}}
\end{table*} 
\section{Results from the observations}
\label{sec:results}
In Table\,2, we present both total and polarised flux 
densities derived by analysing 
the results of the observations completed with the Effelsberg 100-m 
radio telescope at the four frequencies plus those extracted from the NVSS.
These measurements plus those derived from the literature are shown in 
Appendix 1.
\subsection{Percentage of polarised emission}
\label{subsec:m}

The sample constructed by Fanti et al. (1990) contains 47 sources.
We considered them to be polarised when the intensity of the polarised 
emission is 3 times the rms error estimated source by source for the polarised
emission. This provides in general lower limits of 3\% at 2.64\,GHz, 2\% at 
4.85\,GHz, 1\% at 8.35\,GHz,  and 1\% at 10.45\,GHz.
Table\,3 summarizes the values of the polarisation parameters for
sources with polarised flux densities above the detection limits. 

A high fraction of sources have polarised emission, if any, below the
detection limits of our Effelsberg observations. 
We found that 22 of them are polarised at 10.45\,GHz,
16 sources are polarised at 8.35\,GHz,
10 sources are polarised at 4.85\,GHz, and
10 sources are polarised at 2.64\,GHz.
The number of sources with detected polarised emission clearly
decreases from high to low frequency. The percentage of polarised emission 
$m$ ranges between $\sim$6\% and 
$\sim$12\% for the most polarised sources 3C138, 3C286, and 3C454. 

Because most of the sources in the sample have polarised
emission below the detection limits of our observations,
the median values of $m$ for the entire sample are below 3\% at 2.64\,GHz,
below 2\% at 4.85\,GHz, below 1\% at both 8.35\,GHz and 10.45\,GHz, and
below 1.58\% at 1.4\,GHz. 
\tabcolsep0.1cm
\begin{table*}[h]
\centering
\caption{Percentage of polarised flux density and position angle of the 
electric vector at the five frequencies. }
\tiny
\begin{tabular}{llrrrrrrrrrrr}
\hline
Name  &        & $LS$ &$m_{1.4}$ & $m_{2.64}$ & $m_{4.85}$ & $m_{8.35}$ & $m_{10.45}$ & $\chi_{1.4}$ & $\chi_{2.64}$ & $\chi_{4.85}$ & $\chi_{8.35}$ & $\chi_{10.45}$  \\
      &        & [kpc]       &    [$\%$]    &    [$\%$]    &      [$\%$]  &   [$\%$]    &        [$\%$]   & [deg]              & [deg]              & [deg]             & [deg]             & [deg]      \\
\hline
3C43  &0127+23 & 12.85 &      &     &     & 2.1$\pm$0.6 & 2.1$\pm$0.4 &     &          &           & 2$\pm$2   & 13$\pm$3   \\
3C48  &0134+32 & 4.00 & 0.44$\pm$0.12 &     & 3.4$\pm$0.8 & 4.9$\pm$0.6 & 5.6$\pm$0.4 & --70.8$\pm$0.1&          & --78$\pm$2 & --68$\pm$2 & --61$\pm$2  \\
3C49  &0138+13 & 3.85 &      &     &     &     &     &               &          &           &           &            \\
3C67  &0221+27 & 6.45 & 0.56$\pm$0.12 &     &     & 3.9$\pm$0.6 & 3.7$\pm$0.5 & 59.3$\pm$0.5 &          &           & 62$\pm$2  & 62$\pm$3   \\
4C34.07&0223+34& 4.04* &      &     &     &     &     &             &          &           &           &            \\
4C16.09&0316+16& 1.3$^+$ &      &     &     &     &     &          &          &           &           &            \\
       &0319+12& 0.2$^+$ &3.80$\pm$0.18 & 5.6$\pm$0.7 & 6.3$\pm$0.8 & 5.8$\pm$0.6 & 5.0$\pm$0.4 & 41.0$\pm$0.1 & 59$\pm$3 & 73$\pm$2  & 74$\pm$2  & 71$\pm$2   \\
3C93.1&0345+33 & 0.7 &    &     &     &     & 2.4$\pm$0.4 &           &          &           &           & 59$\pm$3   \\ 
4C76.03&0404+76& 0.6$^+$&       &     &     &     &     &             &          &           &           &            \\ 
OF247  &0428+20& 0.49 &      &     &     &     &     &           &          &           &           &            \\
3C119& 0429+41 & 0.9 &      &     &     & 5.9$\pm$0.6 & 8.8$\pm$0.4 &       &          &           & 49$\pm$2  & 179$\pm$2  \\
3C138& 0518+16 & 3.25 & 7.2$\pm$0.25 &10.3$\pm$0.7 & 9.9$\pm$0.8 &11.7$\pm$0.7 & 9.9$\pm$0.4 & --11.1$\pm$0.0& --11$\pm$2& --13$\pm$2 & --11$\pm$2 & --7$\pm$2   \\
3C147& 0538+49 & 2.58 &      &     &     &     & 1.2$\pm$0.4 &          &          &           &           & 189$\pm$2  \\ 
3C186& 0740+38 & 5.14 &      &     &     &     &     &          &          &           &           &            \\ 
3C190& 0758+14 & 11.19 &     &     &     &     &     &          &          &           &           &            \\ 
3C216& 0906+43 & 33.44 &     &     &     &     &     &          &          &           &           &            \\ 
3C237& 1005+07 & 7.11 &      &     &     &     & 1.2$\pm$0.4 &             &          &           &           & 18$\pm$4   \\ 
3C241& 1019+22 & 3.82 &     &     &     &     &     &           &          &           &           &            \\ 
4C31.38&1153+31& 3.96 &     &     &     &     & 1.9$\pm$0.4 &             &          &           &           & 16$\pm$4   \\ 
3C268.3&1203+64& 5.28 &     &     &     & 2.8$\pm$0.6 & 3.0$\pm$0.4 &      &          &           & --97$\pm$2 & -98$\pm$6  \\ 
ON343 &1225+36& 0.23$^+$&     &     &     &     &     &           &          &           &           &            \\ 
3C277.1&1250+56& 4.57 &     &     &        & 3.0$\pm$0.6 & 2.6$\pm$0.5 &  &          &          & 11$\pm$2  & 7$\pm$6    \\ 
4C32.44&1323+32& 0.18$^+$ &      &     &     &     &     &         &          &           &           &            \\ 
3C286& 1328+30 & 15.81 & 6.70$\pm$0.23 &10.9$\pm$0.7 &10.8$\pm$0.8 &11.7$\pm$0.7 &11.7$\pm$0.5 & 35.0$\pm$0.0 & 31$\pm$2 & 33$\pm$2  & 33$\pm$2  & 33$\pm$2   \\ 
3C287& 1328+25 & 0.43 & 0.65$\pm$0.12 & 3.6$\pm$0.7 & 3.7$\pm$0.8 & 3.2$\pm$0.6 & 3.3$\pm$0.4 & 29.9$\pm$0.2 & 105$\pm$2&--32$\pm$2  & --15$\pm$2 & --6$\pm$2   \\
4C62.22&1358+62& 0.17$^+$ &     &     &     &     &     &         &          &           &           &            \\
       &1413+34& 0.26 &          & 3.1$\pm$0.7 &     &     &     &      & 199$\pm$3&           &           &            \\   
3C298&1416+06  & 9.1 &    &     &     & 1.8$\pm$0.6 & 2.1$\pm$0.4 &         &          &           & --42$\pm$3 & --30$\pm$4  \\ 
3C299&1419+41  & 36.31 &     &     &     &     &     &         &          &           &           &            \\   
OQ172  &1442+10  & 0.1$^+$ & 1.30$\pm$0.13 & 4.5$\pm$0.9 & 2.6$\pm$0.8 & 2.2$\pm$0.6 &     & --86.2$\pm$0.3& 212$\pm$4& 69$\pm$2  & 76$\pm$3  &            \\
3C303.1&1443+77& 1.28 &     &     &     &     &     &         &          &           &           &            \\ 
3C305.1&1447+77& 7.86 &     &     &     &     &     &         &          &           &           &            \\
3C309.1&1458+71& 9.25 & 1.20$\pm$0.13 & 3.7$\pm$0.7 & 2.4$\pm$0.7 & 2.7$\pm$0.6 & 2.5$\pm$0.4 & 39.9$\pm$0.1 & 104$\pm$2&252$\pm$2  & 247$\pm$2 & 243$\pm$2  \\
3C318& 1517+20 & 3.25 &     &     & 3.4$\pm$0.8 & 6.6$\pm$0.6 & 9.2$\pm$0.6 &   &          & 60$\pm$2  & --14$\pm$2 & --24$\pm$2  \\
     & 1600+33 & $<$0.4 &    &     &     &     &     &       &          &           &           &            \\
OS111& 1607+26 & 0.19$^+$&     &     &     &     &     &      &          &           &           &            \\   
3C343&1634+62  & 11.48 &      &     &     &     & 1.4$\pm$0.4 &        &          &           &           & --104$\pm$3 \\   
3C343.1&1637+62& 1.54 &      &     &     &     &     &       &          &           &           &            \\
3C346&1641+17  & 27.01 & 2.20$\pm$0.14 &     &     &     & 1.8$\pm$0.4 & --82.7$\pm$0.1&          &           &           & --4$\pm$3   \\  
4C39.56&1819+39& 3.20 &      &     &     &     &     &       &          &           &           &            \\
3C380& 1828+48 & 55.62 & 0.45$\pm$0.12 &     &     &     &     & 44.2$\pm$0.1 &          &           &           &            \\  
4C29.56 &1829+29& 12.40 &     &     &     &     &     &        &          &           &           &            \\
4C11.69&2230+11& 9.7 & 1.6$\pm$0.13 & 4.0$\pm$0.7 & 3.6$\pm$0.8 & 3.7$\pm$0.6 & 3.6$\pm$0.4 & --61.1$\pm$0.1& 7$\pm$2  & 48$\pm$2  & 58$\pm$2  & 60$\pm$2   \\ 
3C454 &2249+18 & 8.39 & 3.6$\pm$0.16  & 5.8$\pm$0.7 & 9.3$\pm$0.8 &11.2$\pm$0.7 &11.2$\pm$0.4 & 53.4$\pm$0.1 & 41$\pm$2 & 87$\pm$2  & 98$\pm$2  & 101$\pm$2  \\ 
3C454.1&2248+71& 5.5 &      &     &     &     &     &         &          &           &           &            \\ 
3C455&2252+12  & 11.75 & 2.10$\pm$0.14 & 3.4$\pm$0.7 &     &     & 1.5$\pm$0.5 & --7.5$\pm$0.2 & 149$\pm$3&           &           & 151$\pm$8  \\   
     &2342+82  & 0.72$^+$ &     &     &     &     &     &        &          &           &           &            \\  
\hline
\normalfont
\end{tabular}
\normalfont
\smallskip\noindent
\flushleft{\normalsize {Table\,3 is organized
as follows: column 1, source name; column 2, other name; column 3, source 
linear size. Asterisks mark sources for which new redshifts have been obtained
and new linear sizes were calculated;  $^+$ signs mark sources with
improved measurements of their angular sizes (Rossetti et al. 2005; 0319+12
Mantovani in prep.; 1442+10 Udomprasert et al. 1997);
columns 4 to 8, percentage of polarised emission $m$; 
columns 9 to 13, position angle of the electric vector $\chi$.}
}
\end{table*}
\subsection{Depolarisation index and rotation measure}
\label{subsec:RM}
 
Table 4 summarizes the values of the depolarisation index
and of the observed RM (RM$_{obs}$) and source rest frame RM  
(RM$_{rf}$ = RM$_{obs} \times (1+z)^2$) both given in rad m$^{-2}$. 
The depolarisation index DP = $m_l / m_h$, is
defined to be the ratio of the percentages of polarised emission at the lower 
($m_l$) to the higher ($m_h$) frequency.
The present observations complemented with those taken from Tabara \& Inoue 
(1980), allow us to determine the values of the RMs
for 16 sources in the list. The observing
frequencies of the Effelsberg plus NVSS observations are suitably 
separated for a proper determination of 
the RMs. In particular, they allow us to apply unambiguous $n\pi$ de-rotation 
to the observed polarisation E-vector position angle $\chi$ at the 
various frequencies.  When applied, these rotations always yield a linear 
regression with a least squares fit very close to 1, expect for an optimal 
fit, for the $\lambda^2$ rotation.
We also note that there is no significant difference between the values 
derived from our Effelsberg observations and those from Tabara \& 
Inoue (1980), which were acquired about 30 years earlier.
\tabcolsep0.1cm
\begin{table*}[h]
\centering
\caption{Depolarisation indices and RMs.}
\tiny
\begin{tabular}{lllcrrrrrrrrr}
\hline
Name     &        &  z   & Id& DP              & DP             & DP             & DP            & RM$_{obs}$     &  lm-1  & RM$_{obs+TI}$  &  lm-2  & RM$_{rf}$ \\
         &        &      &   & $_{8.35/10.45}$ & $_{4.85/8.35}$ & $_{2.64/4.85}$ & $_{1.4/2.64}$ & [rad m$^{-2}]$ &        & [rad m$^{-2}$] &        & [rad m$^{-2}$]\\
\hline
3C43     &0127+23 & 1.46 & Q & 1.00            &                &               &                &                &        &  $-$69         & 0.9965 & $-$419 \\
3C48     &0134+32 & 0.37 & Q & 0.87            &     0.69       &               &                & $-$79          & 0.9994 & $-$64          & 0.9900 & $-$120 \\
3C49     &0138+13 & 0.62 & G &                 &                &               &                &                &        &                &        &          \\
3C67     &0221+27 & 0.31 & G &            1.05 &                &               &                & $-$1           & 0.9999 & $-$67          & 0.9945 & $-$115 \\
4C34.07  &0223+34 & 2.91*& Q &                 &                &               &                &                &        &                &        &          \\
4C16.09  &0316+16 & 0.97*& Q &                 &                &               &                &                &        &                &        & $-$249 \\
         &0319+12 & 2.67 & Q &            1.16 &           1.09 & 0.89          & 0.07           & $-$12          & 0.9521 & $-$19          & 0.7698 & $-$164 \\
3C93.1   &0345+33 & 0.24 & G &                 &                &               &                &                &        &                &        &          \\ 
4C76.03  &0404+76 & 0.6  & G &                 &                &               &                &                &        &                &        &          \\ 
OF247    &0428+20 & 0.22 & G &                 &                &               &                &                &        &                &        &          \\
3C119    & 0429+41& 1.023& Q &            0.66 &                &               &                &                &        & 1928           & 0.9663 & 3900   \\
3C138    &0518+165& 0.76 & Q &            1.18 &           0.85 &        1.04   &           0.70 & $-$1           &  0.04    & $-$6         & 0.8786 & $-$19  \\
3C147    &0538+49 & 0.55 & Q &                 &                &               &                &                &        &                &        &          \\ 
3C186    &0740+38 & 1.06 & Q &                 &                &               &                &                &        &                &        &          \\ 
3C190    &0758+14 & 1.2  & Q &                 &                &               &                &                &        &                &        &          \\ 
3C216    &0906+43 & 0.67 & Q &                 &                &               &                &                &        &                &        &          \\ 
3C237    &1005+07 & 0.88 & G &                 &                &               &                &                &        &                &        &          \\ 
3C241    &1019+22 & 1.62 & G &                 &                &               &                &                &        &                &        &          \\ 
4C31.38  &1153+31 & 0.42 & Q &                 &                &               &                &                &        &                &        &          \\ 
3C268.3  &1203+64 & 0.37 & G &            0.93 &                &               &                &                &        &  114           & 0.9853 & 213    \\ 
ON343    &1225+36 &1.973*& Q &                 &                &               &                &                &        &                &        &          \\ 
3C277.1  &1250+56 & 0.32 & Q &            1.15 &                &               &                &                &        & $-$59          & 0.9961 & $-$102 \\ 
4C32.44  &1323+32 & 0.37 & G &                 &                &               &                &                &        &                &        &          \\ 
3C286    &1328+30 & 0.85 & Q &            0.99 &           0.91 &         1.01   & 0.61          &                &        &                &        &          \\ 
3C287    &1328+25 & 1.06 & Q &            0.97 &           1.16 &        0.97   & 0.18           & $-$148         & 0.9957 &   $-$146       & 0.9954 & $-$618 \\
4C62.22  &1358+62 & 0.43 & G &                 &                &               & 0.02           &                &        &                &        &          \\
         &1413+34 &      &   &                 &                &               &                &                &        &                &        &          \\   
3C298    &1416+06 & 1.44 & Q &            0.86 &                &               &                &                &        &      96        & 0.9911 &  573   \\ 
3C299    &1419+41 & 0.37 & G &                  &               &               &                &                &        &                &        &          \\   
OQ172    &1442+10 & 3.52 & Q &                  &          1.18 &        1.73   & 0.29           & $-$66          & 0.9994 & $-$71          & 0.9376 & $-$1451  \\
3C303.1  &1443+77 & 0.2  & G &                  &               &               &                &                &        &                &        &          \\ 
3C305.1  &1447+77 & 1.13 & G &                  &               &               &                &                &        &                &        &          \\
3C309.1  &1458+71 & 0.91 & Q &             1.08 &          0.89 &        1.54   & 0.32           & 60             & 0.9996 &     46         & 0.9830 &   169  \\
3C318    &1517+20 & 1.57 & G &             0.72 &          0.52 &               &                &    498         & 0.9987 &    342         & 0.9995 & 2260   \\
         &1600+33 & 1.1* & G?&                  &               &               &                &                &        &                &        &          \\
OS111    &1607+26 & 0.47*& G &                  &               &               &                &                &        &                &        &          \\   
3C343    &1634+62 & 0.99 & Q &                  &               &               &                &                &        &                &        &          \\   
3C343.1  &1637+62 &0.75  & G &                  &               &               &                &                &        &                &        &          \\
3C346    &1641+17 & 0.16 & G &                  &               &               &                &                &        &                &        &          \\  
4C39.56  &1819+39 & 0.80 & G &                  &               &               &                &                &        &                &        &          \\
3C380    &1828+48 & 0.69 & Q &                  &               &               &                &                &        &                &        &          \\  
4C29.56  &1829+29 & 0.84 & G &                  &               &               &                &                &        &                &        &          \\
4C11.69  &2230+11 & 1.04 & Q &             1.03 &          0.97 &        1.10   & 0.40           & $-$39          & 0.9694 &  $-$38         & 0.9894 & $-$157  \\ 
3C454    &2249+18 & 1.76 & Q &             1.02 &               &         0.62  & 0.62           & $-$88          & 1.000  &  $-$88         & 0.9997 & $-$669 \\ 
3C454.1  &2248+71 & 1.84 & G &                  &               &               &                &                &        &                &        &          \\ 
3C455    &2252+12 & 0.54 & G &                  &               &               & 0.62           &   $-$68        & 0.9402 &   82           & 0.9923 &   194  \\   
         &2342+82 & 0.74 & Q &                  &               &               &                &                &        &                &        &          \\  
\hline
\normalfont
\end{tabular}
\normalfont
\smallskip\noindent
\flushleft{\normalsize { Table 4 is organized as follows: 
column 1, source name; column 2, other name; column 3, redshift; 
redshifts marked with ``*'' are taken from the following pubblications:
0223+34 Willott et al. (1998); 0316+16 Labiano et al. (2007);
1225+36 Lawrence et al. (1994); 1600+33 photometric redshift from Stanghellini 
et al. (2005); 1607+26 Stanghellini et al. (1993); column 4, optical 
identification. The source 3C455 has been re-classified as galaxy by 
Hes et al. (1966); columns 5 to 8, Depolarisation Indices DP; 
column 9, Rotation Measure RM derived from our new Effelsberg observations;
column 10, ls-1: the least square fit of the linear regression of the values 
in column 9; column 11, Rotation Measure RM derived from our new Effelsberg 
observations, complemented with data from Tabara \& Inoue (1980);
column 12, ls-2: the least square fit of the linear regression of the RM 
values in column 11; column 13, rest frame RM calculated using the 
RM$_{obs+TI}$ values.}
}
\end{table*}
%
%
%\clearpage
%
%\newpage
%
\section{Discussion}
\label {sec:discussion}

In the following we discuss trends of the depolarisation index,
the achieved values of the rotation measure, and the distribution of 
the percentage of polarised emission versus the linear sizes 
of the observed sources. 
\subsection{General trends} 
The depolarisation index DP has been computed for four
pairs of frequencies. The mean and the median values of the DP are 
reported in Table\,5.
\tabcolsep0.2cm
\begin{table}[h]
\centering
\caption{The mean and the median values of the depolarisation index }
\begin{tabular}{lll}
\hline
Depolarisation index&   DP$_{mean}$ &     DP$_{median}$   \\
\hline
DP$_{8.35/10.45}$   & 0.98$\pm$0.15 & 1.00\,$_{-0.03}\hskip-21pt\,^{+0.02}$  \\
DP$_{4.85/8.35}$    & 0.90$\pm$0.20 & 0.91\,$_{-0.006}\hskip-25pt\,^{+0.18}$ \\
DP$_{2.64/4.85}$    & 1.11$\pm$0.33 & 1.01\,$_{-0.12}\hskip-21pt\,^{+0.09}$ \\
DP$_{1.4/2.64}$     & 0.33$\pm$0.23 & 0.32\,$_{-0.14}\hskip-21pt\,^{+0.30}$ \\
\hline
\end{tabular}
\end{table}
Clearly a drop in the fractional polarisation appears in the radio emission
at frequencies below $\sim$2\,GHz.
The number of polarised sources in our list of CSS sources 
drops from 46\% at 10.45\,GHz to 23\% at 2.64 GHz, and they have a median 
fractional polarisation of below 2\% at 5.0\,GHz that agrees with previous 
findings (for example Saikia et al. 1987).

In Fig.s 2 to 17 (Appendix 1), we plot the fractional polarisation values 
$m$ derived from 
our Effelsberg observations complemented with those extracted from the NVSS 
at 1.4\,GHz and those taken from the  Tabara \& Inoue (1980) catalogue. 
The general trend 
is  a quick decrease in the fractional polarisation with increasing wavelength.
However, $m$ does not always drop to zero at longer wavelengths 
($\lambda > $ 49\,cm) as predicted by the Burn (1966) model. The Burn model 
(dashed line) and the Tribble (1991) model (solid line) have been plotted in 
Figs. 2 to 17, according to Eqs. (1) and (2) in Fanti et al. (2004). 
For both models, we have adopted as the intrinsic fractional polarisation 
$m_0$ the
one measured by us at 10.45\,GHz. For the Tribble model, which assumes RM 
randomly distributed and a distribution of cell sizes, the ratio of the 
characteristic scale representing the largest cell scale to the observing 
beam is taken to be equal to 1. With these assumptions, the model corresponds 
to the 
highest possible polarisation at long wavelengths, thus an upper limit to the 
expected fractional polarisation behaviour. 

Very low fractional polarisation for sources belonging to this class is also
found with interferometric observations. For example,  
Peck \& Taylor (2000) did not detect any linear polarisation with VLBI
observations at 8.35 GHz on a sample of 21 Compact Symmetric Objects (CSOs are
CSS sources that exhibit core and lobe emission on each side of the core).
While this strengthens the assumption that beam depolarisation plays a minor
r\^ole it also means that the polarisation characteristics of CSS sources
are not yet fully clarified.
%
%\clearpage
%\newpage
%
%\begin{figure}[!hb]
%\addtocounter{figure}{+0}
%\centering
%%\resizebox{8cm}{!}{\rotatebox{270}{\includegraphics{plotls14.ps}}}
%\includegraphics[width=8cm]{plotls14.ps}
%\caption{Percentage of polarised emission versus source linear sizes at
%1.4\,GHz. Dots are quasars, and open circles are galaxies.}
%\end{figure}
%\vspace{-0.3cm}
%
%
%\begin{figure}[!hb]
%\addtocounter{figure}{+0}
%\centering
%%\resizebox{8cm}{!}{\rotatebox{270}{\includegraphics{plotls27.ps}}}
%\includegraphics[width=8cm]{plotls27.ps}
%\caption{Percentage of polarised emission versus source linear sizes at
%2.64\,GHz. Dots are Quasars; open circles are Galaxies.}
%\end{figure}
%\vspace{-0.3cm}
%
%\clearpage
%
%\begin{figure}[!hb]
%\addtocounter{figure}{+0}
%\centering
%%\resizebox{8cm}{!}{\rotatebox{270}{\includegraphics{plotls50.ps}}}
%\includegraphics[width=8cm]{plotls50.ps}
%\caption{Percentage of polarised emission versus source linear sizes at
%4.85\,GHz. Dots are Quasars; open circles are Galaxies.}
%\end{figure}
%\vspace{-0.3cm}
%
\begin{figure}[!hb]
\addtocounter{figure}{+0}
\centering
%\resizebox{8cm}{!}{\rotatebox{270}{\includegraphics{plotls84.ps}}}
\includegraphics[width=8cm]{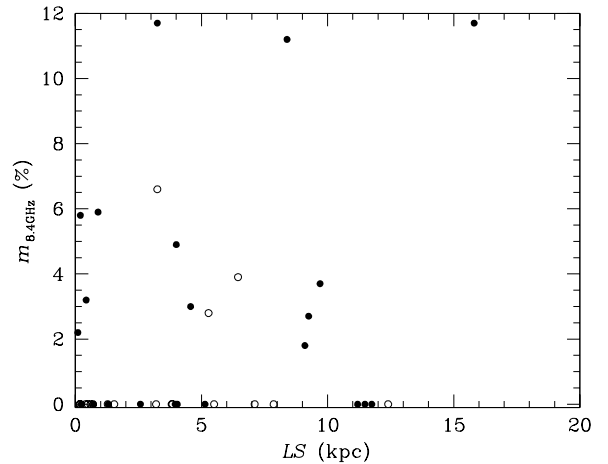}
\caption{Percentage of polarised emission versus source linear sizes at
8.35\,GHz. Dots are Quasars; open circles are Galaxies.}
\end{figure}
\vspace{-0.3cm}
%
%\clearpage
%
%\begin{figure}[!hb]
%\addtocounter{figure}{+0}
%\centering
%%\resizebox{8cm}{!}{\rotatebox{270}{\includegraphics{plotls11.ps}}}
%\includegraphics[width=8cm]{plotls11.ps}
%\caption{Percentage of polarised emission versus source linear sizes at
%10.45\,GHz. Dots are Quasars; open circles are Galaxies.}
%\end{figure}
%\vspace{-0.3cm}
%
%
\subsection{Model fitting}
Rotation Measures were calculated for about 1/3 of the sources in our list 
using use of our own measurements, NVSS and Tabara \& Inoue data. The 
observed RMs were calculated for 16 sources with at least two measurements 
of the position angles achieved by Effelsberg and NVSS observations.
Plots of the position angle of the electric vector in degrees versus 
the $\lambda^2$ in m$^2$ are shown in Figs.\,2--17. 

In all cases, the $\lambda^2$ law is closely followed for least squares 
estimates of the linear regression close to 1. To search for possible 
variability,
 we determined the RM both from our new measurements only and from all 
available measurements, including those of Tabara \& Inoue, which were 
measured about 30 years ago. No significant difference was found. 

The observing
frequencies were suitably separated to enable an unambiguous rotation of
the polarisation E-vectors. By analysing the combination
of RM and fractional polarisation, we measured $\lambda^2$ rotation and 
depolarisation for all sources, which corresponded to depolarisation produced 
inside the 
source, and rotation produced outside. However, since $\delta\chi>\pi$,  we
 actually observed not mixed-in gas but a foreground screen. 
A revised Tribble
model such as that proposed by Rossetti et al. (2008) with a 
``partial coverage'' 
of the source of radio emission by NLRs may account for the depolarisation 
behaviour.

Comments are necessary in a few cases.

\noindent
- 3C67: the RM values calculated making use of the 
three position angles of Table\,3 give an excellent least squares fit 
the by means of linear regression after rotating by both 
$+ \rm and - \pi$ the nominal value of $59^\circ.3$
obtained at 1.4\,GHz, namely 69 rad m$^{-2}$ and --71 rad m$^{-2}$.
The last value is close to that ($-$67 rad m$^{-2}$) calculated also using the 
RMs listed in the Tabara \& Inoue catalogue. 
3C138: the RM value calculated with the position angles  listed in Table\,3 
is --1 rad m$^{-2}$ with a very poor mean least squares fit. 

\noindent
- 3C286: the RM 
value is not given as the position angle of the electric vector (33$^\circ$) 
for this source is taken as a reference to calibrate the position angles 
at all the observing frequencies. 

\noindent
- 3C455: the least squares fit obtained with 
the values of Table\,3 is worse than that achieved 
using all the available data. 

In conclusion, we adopt in all cases the RMs listed in Col.\,11 of Table\,4 
calculated using the present measurements of the position angles plus those 
taken from the catalogue by Tabara \& Inoue. 
The RMs in the source rest frames range between --20 rad m$^{-2}$ for 3C138 
and $\sim$3900 rad m$^{-2}$ for 3C119. Seven sources have a 
RM $>$400 rad m$^{-2}$.

Examples of CSS sources with high values of $\mid$RM$\mid$ ($\geq$ 1000 rad
m$^{-2}$ in the source rest frame) have been found in the past
(O'Dea 1998 and references therein).

Three of them are in our list with measured integrated RMs, namely 3C119, 
1442+101 and 3C318 for which we measured 3900 rad m$^{-2}$, --1450 
rad m$^{-2}$, and 2260 rad m$^{-2}$ respectively. 
For the source 1442+101 (OQ172), observed by Udomprasert (1998) with 
the VLBA, RMs up to 22400 rad m$^{-2}$ were found. For 3C119, Kato et al. 
(1987)
found 3400 rad m$^{-2}$ using single dish measurements. For 3C318,  Taylor et 
al. (1992) determined a RM of 1400 rad m$^{-2}$ using VLA observations.  
There are however CSS sources in our sample with smaller values of RM.

As described in more detail in Appendix 2, the source 3C298 shows evidence 
of temporal variability in the fractional polarisation.

\subsection{The ``Cotton Effect''}
As an example, we show the percentage of polarised emission versus the 
source's linear size in kpc at a frequency of 8.35\,GHz in Fig.\,1.
Linear dimensions for sources labelled with ``*'' in Table\,4, for which $z$ 
became available after 1990, were calculated.
We note that in the plot sources with linear sizes greater than 15\,kpc are
omitted and sources with polarised emission below the detection
limits are plotted assuming $m =0 $.

The most unexpected result that we find is the distribution of points drawn in
the plots at the Effelsberg and the NVSS observing frequencies. Polarised 
sources are distributed all over the $m$ -- $LS$ space. This seems to clash 
with the findings of Cotton et al. (2003) and Fanti et al. (2004) for the 
B3-VLA sample of CSS sources. 
Cotton et al. (2003) making use of NVSS at 1.4\,GHz found that sources 
smaller than 6\,kpc are weakly polarised, and that polarised sources 
have linear sizes greater than 6\,kpc (``Cotton effect''). 
Both the jet and the counter-jet are included in the source
linear size. This implies a drastic change in the interstellar medium at
about 3\,kpc.
This result was later confirmed by Fanti et al. (2004), who observed the same 
sample with the VLA at 4.9\,GHz and 8.5\,GHz and with the WSRT at 2.64\,GHz 
(Rossetti et al. 2008).

The main difference between the two samples, i.e., the B3-VLA and the present
3CR\&PW sample of CSS sources, is represented by the number of quasars,  
5-10\% and 49\%, respectively. 
Excluding the quasars, we find that the Cotton effect is clearly evident 
again. 
However, we also note three outliers, the galaxies 3C93.1, 3C268.3, and 3C318. 
On the other hand, it has been found by Rossetti et al. (2008) that the Cotton 
effect is not obeyed by all sources. This indicates that the current 
depolarisation scenarios might not fully explain the observed behaviour.
This is also further indication 
that CSS sources optically identified with quasars may represent a separate 
class of objects. 
\section{Conclusions}
\label{sec:conclusions}

With the present observations of a complete sample of CSS sources at 4 
different 
frequencies made with the Effelsberg 100-m radio telescope and from
archived NVSS data, the following results have been reached:

a) It is confirmed that CSS sources are weakly polarised, 
low values of the median fractional polarisation being found. Where $m$ 
can be compared with those obtained by Klein et al. (2003) for the steep 
spectrum extended radio sources selected in the B3-VLA sample, 
larger values are found, which are 2.2\% at 1.4~GHz, 
3.7\% at 2.64\,GHz, 5.2\% at 4.85\,GHz, and 5.8\% at 10.45\,GHz.
The more sensitive VLA observations at 1.4\,GHz of the CSS sources in the 
3CR\&PW sample extracted from the NVSS also show a median value for
the fractional polarisation of $\sim$1.6\%, which is rather low.
 
b) As expected, the percentage of polarised sources decreases from 
higher to lower frequencies. We found 22 sources polarised at 10.45\,GHz,
16 sources polarised at 8.35\,GHz, 10 sources polarised at both 4.85\,GHz
and 2.64\,GHz, and 13 at 1.4\,GHz. In general, the percentage of polarised 
emission remains almost constant
down to 2.64\,GHz, then drastically drops at frequencies below 
$\sim$2\,GHz. There may be cases of repolarisation. Of particular interest, 
if this finding is confirmed, is the case of 3C455.

c) A case of variability with time of the fractional polarisation is
suggested for the source 3C298.

d) The data obtained with these observations allowed us to determine the 
RMs for 16 sources. The RMs in the source rest frames range 
between --20 rad m$^{-2}$ for 3C138 and $\sim$3900 rad m$^{-2}$ for 3C119.
Seven sources show a RM $>$400 rad m$^{-2}$, 
confirming the high values of the RM usually found for CSS sources. 

e) In all cases the $\lambda^2$ law is followed closely. 
The observing
frequencies were suitably separated to enable an unambiguous rotation of
the polarisation E-vectors. Analysing the combination
of RM and fractional polarisation, we identified $\lambda^2$ rotation and 
depolarisation for all sources.
A revised Tribble 
model such as that proposed by Rossetti et al. (2008) with a ``partial 
coverage'' 
of the source of radio emission by NLRs may account for the depolarisation 
behaviour.

f) The $m$ - $LS$ diagram contains sources
polarised at any linear size and at all available frequencies, 
including the NVSS 1.4\,GHz data, i.e., also for sources smaller than 6 kpc
in linear size, in contrast to the so-called ``Cotton effect'' later confirmed 
by Fanti et al. (2004) for the B3-VLA sample of CSS sources. However,
plotting $m$ versus $LS$ for objects optically identified 
with galaxies, the so-called ``Cotton effect'' is reproduced. 
Discussing the interesting, large difference in the
percentage of objects identified as quasars in the two samples was not
a subject of the present investigation.
 
g) CSS sources optically identified with quasars may represent a separate 
class of objects. 

\begin{acknowledgements}

This work is based on observations with the 100-m telescope of the 
MPIfR (Max-Planck-Institut f\"ur Radioastronomie) at Effelsberg.
It has benefited from research funding from the European Community's 
sixth Framework Programme under RadioNet R113CT 2003 5058187.
FM likes to thank Prof. Anton Zensus, Director, for the kind 
hospitality at the Max-Planck-Institut f\"ur Radioastronomie, Bonn, for a 
period during which part of this work was done. 
%Daniela Vergani is warmly thanked for her help with the plots.
The National Radio Astronomy Observatory is a facility of the National Science 
Foundation operated under cooperative agreement by Associated Universities, 
Inc.

%We are very grateful to an anonymous referee for very helpful comments
%and suggestions and for a careful reading of the manuscript of this paper.

\end{acknowledgements}

\clearpage
\newpage
\vspace{3cm}
{\bf Appendix 1}

In this Appendix, we present plots of the fractional polarisation
$m$ derived form our Effelsberg observations complemented with those extracted
from the NVSS at 1.4\,GHz and those taken from Tabara \& Inoue (1980). 
The values of $m$ predicted by the models of Burn and 
Tribble are shown for comparison. Also plotted are polarisation 
angles versus $\lambda^2$ with their corresponding linear best fit.
\begin{figure*}[!hb]
\addtocounter{figure}{+0}
\centering
\includegraphics[width=8cm]{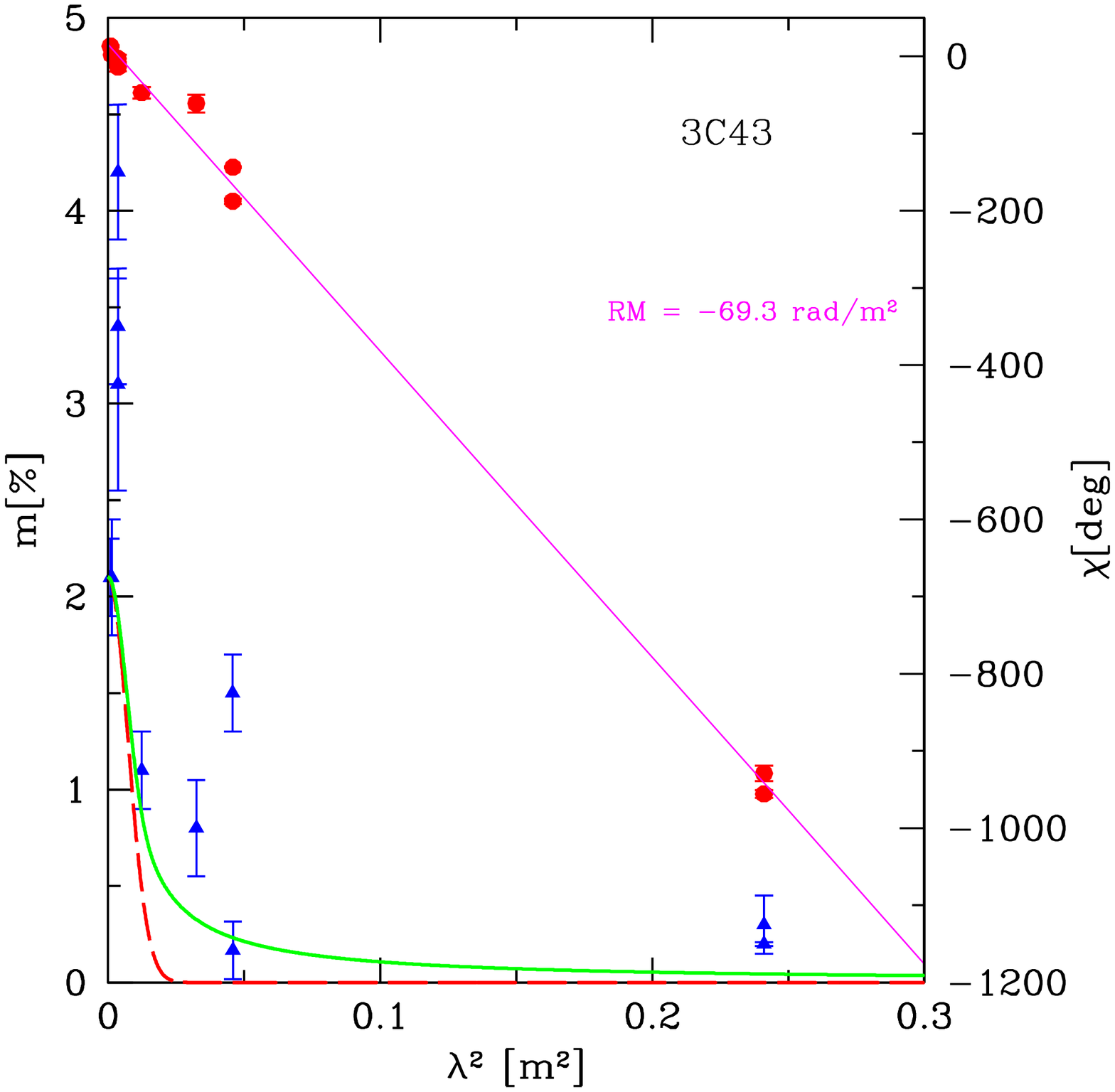}
\includegraphics[width=8cm]{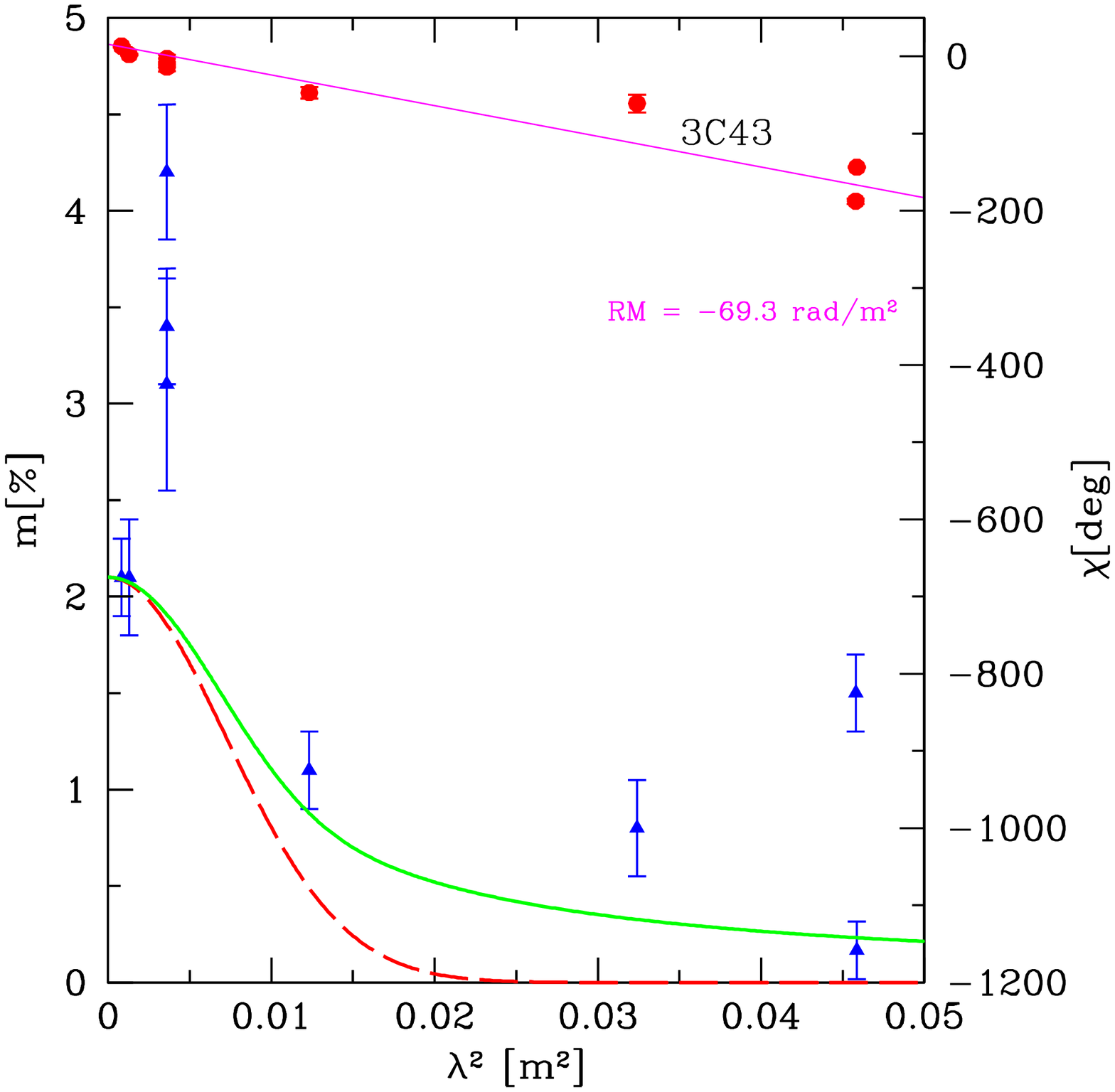}
\caption{Position angles of the electric field vector $\chi$ in deg (dots) and fractional 
polarisation $m$ in \% (triangles) $versus$ $\lambda^2$ in m$^2$ for the source 3C43,
for the full range ({\it left}), and for a narrow range ({\it right}) 
of wavelengths. The solid line represents the Tribble model, the dashed line
the Burn model.}
\end{figure*}
\vspace{-0.3cm}
\begin{figure*}[!hb]
\addtocounter{figure}{+0}
\centering
\includegraphics[width=8cm]{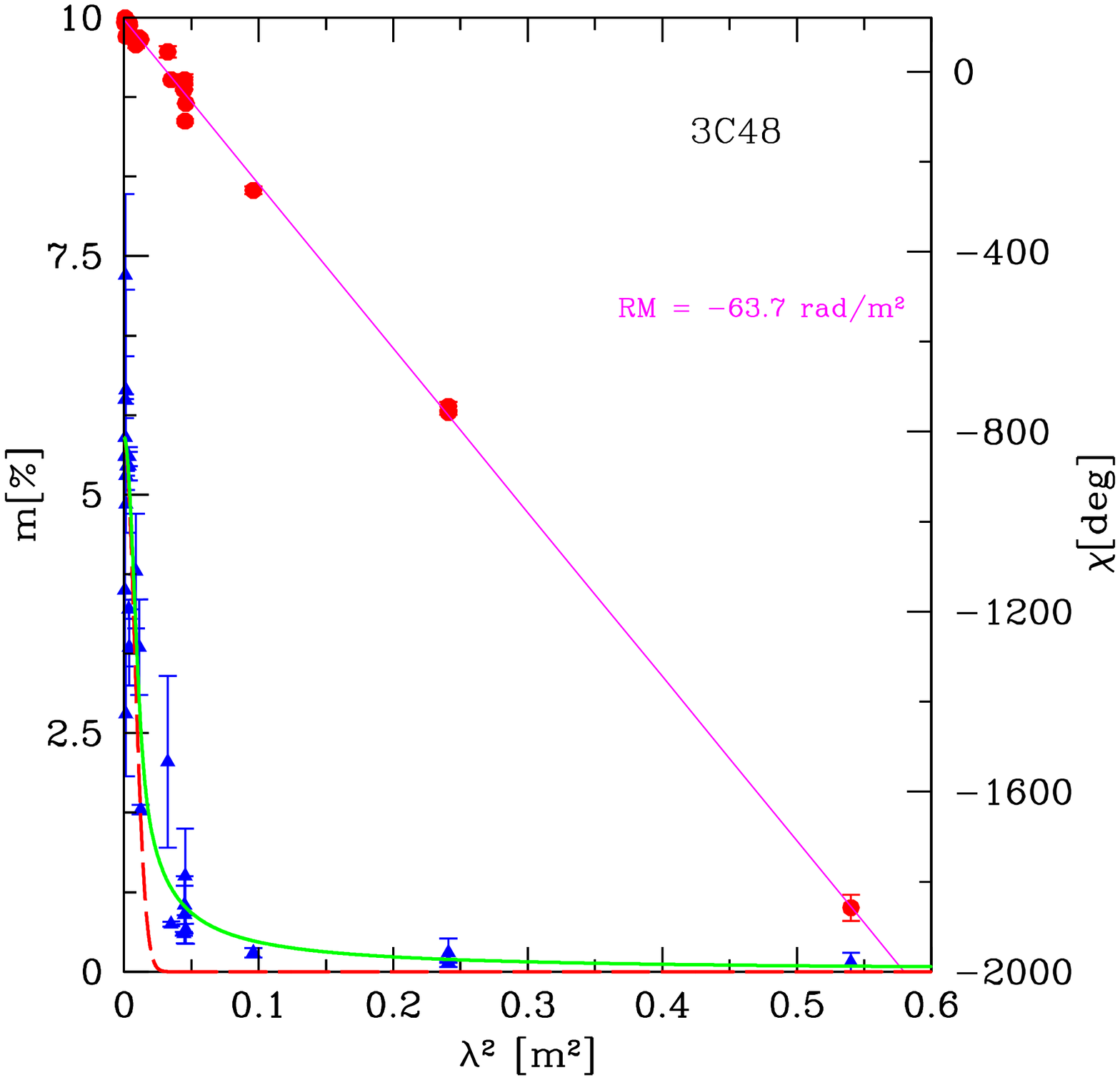}
\includegraphics[width=8cm]{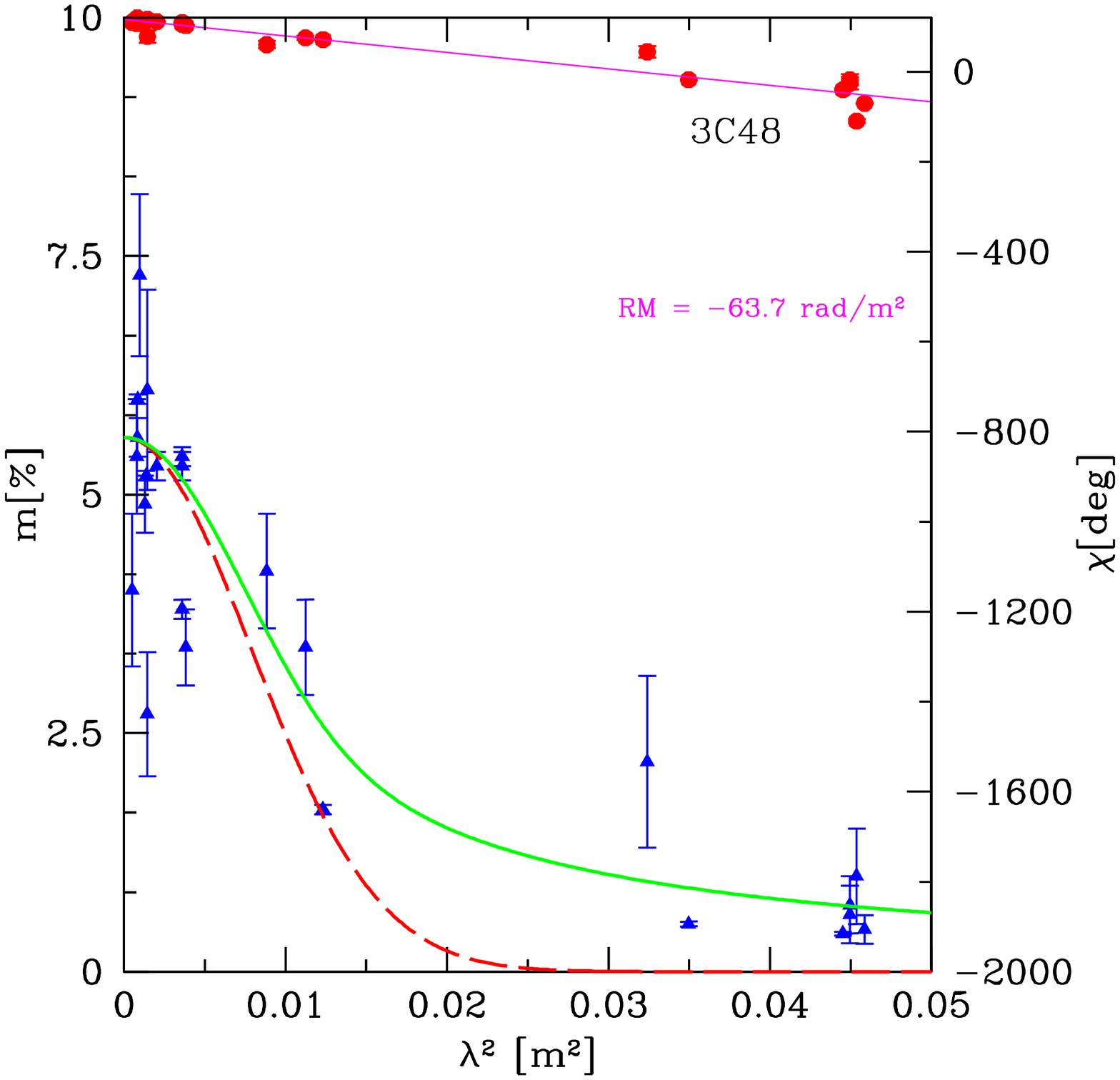}
\caption{Position angles $\chi$ and fractional 
polarisation $m$ for the source 3C48. Layout as in Fig.\,2.}
\end{figure*}
\vspace{-0.3cm}
\begin{figure*}[!hb]
\addtocounter{figure}{+0}
\centering
\includegraphics[width=8cm]{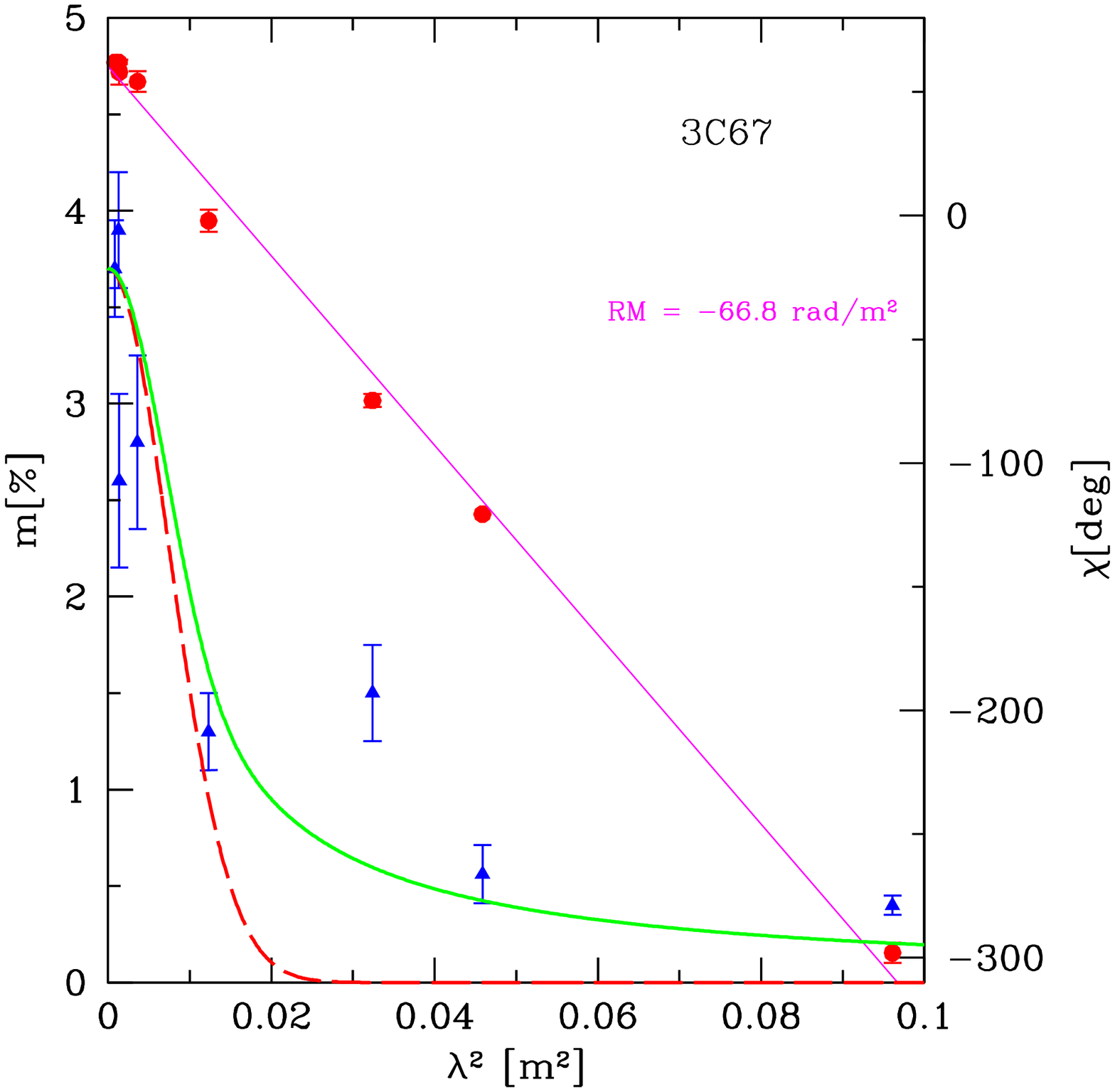}
\caption{Position angles $\chi$ and fractional 
polarisation $m$ for the source 3C67. Layout as in Fig.\,2.}
\end{figure*}
\vspace{-0.3cm}
\begin{figure*}[!hb]
\addtocounter{figure}{+0}
\centering
\includegraphics[width=8cm]{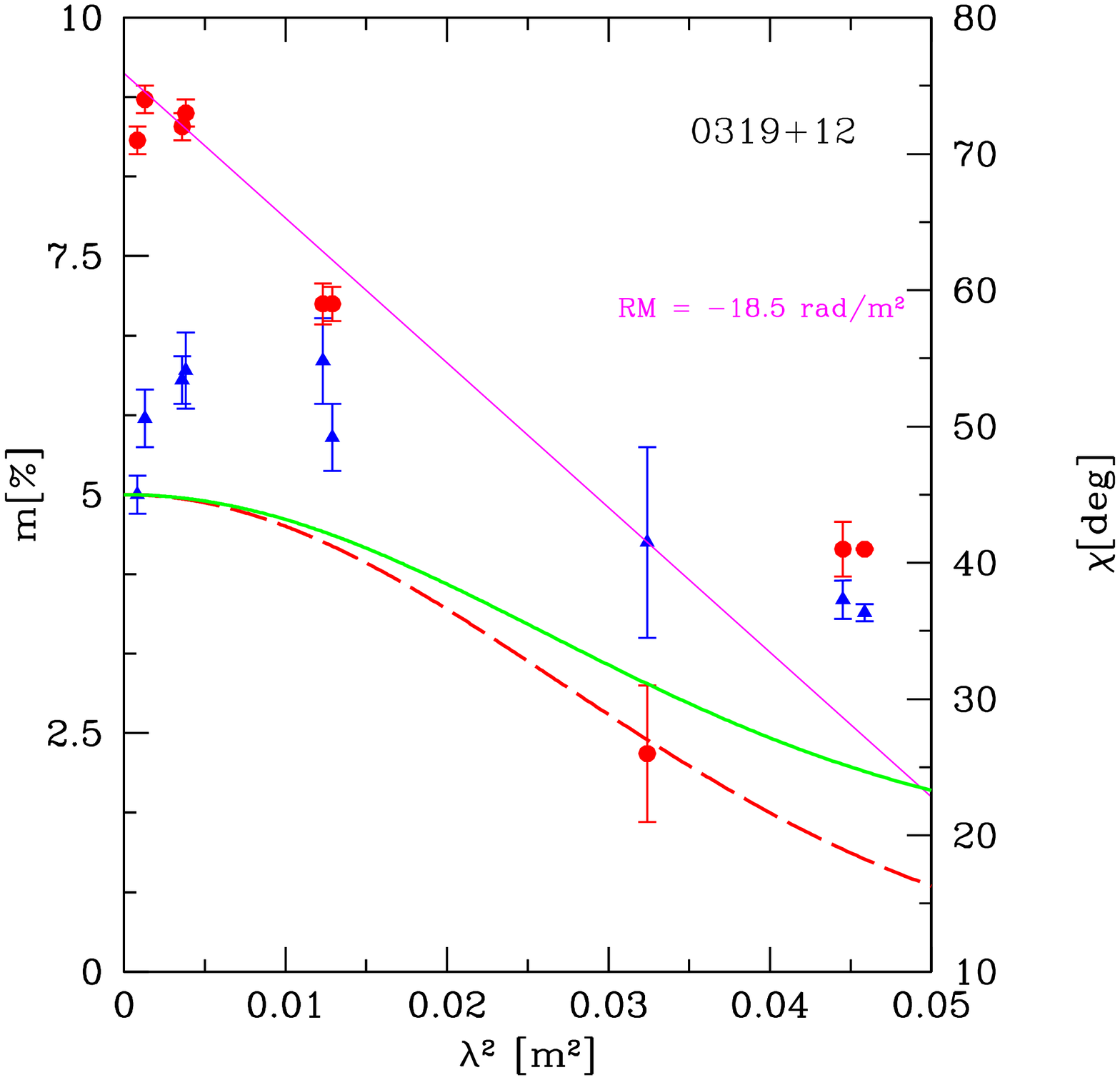}
\caption{Position angles $\chi$ and fractional 
polarisation $m$ for the source 0319+12. Layout as in Fig.\,2.}
\end{figure*}
\vspace{-0.3cm}
\begin{figure*}[!hb]
\addtocounter{figure}{+0}
\centering
\includegraphics[width=8cm]{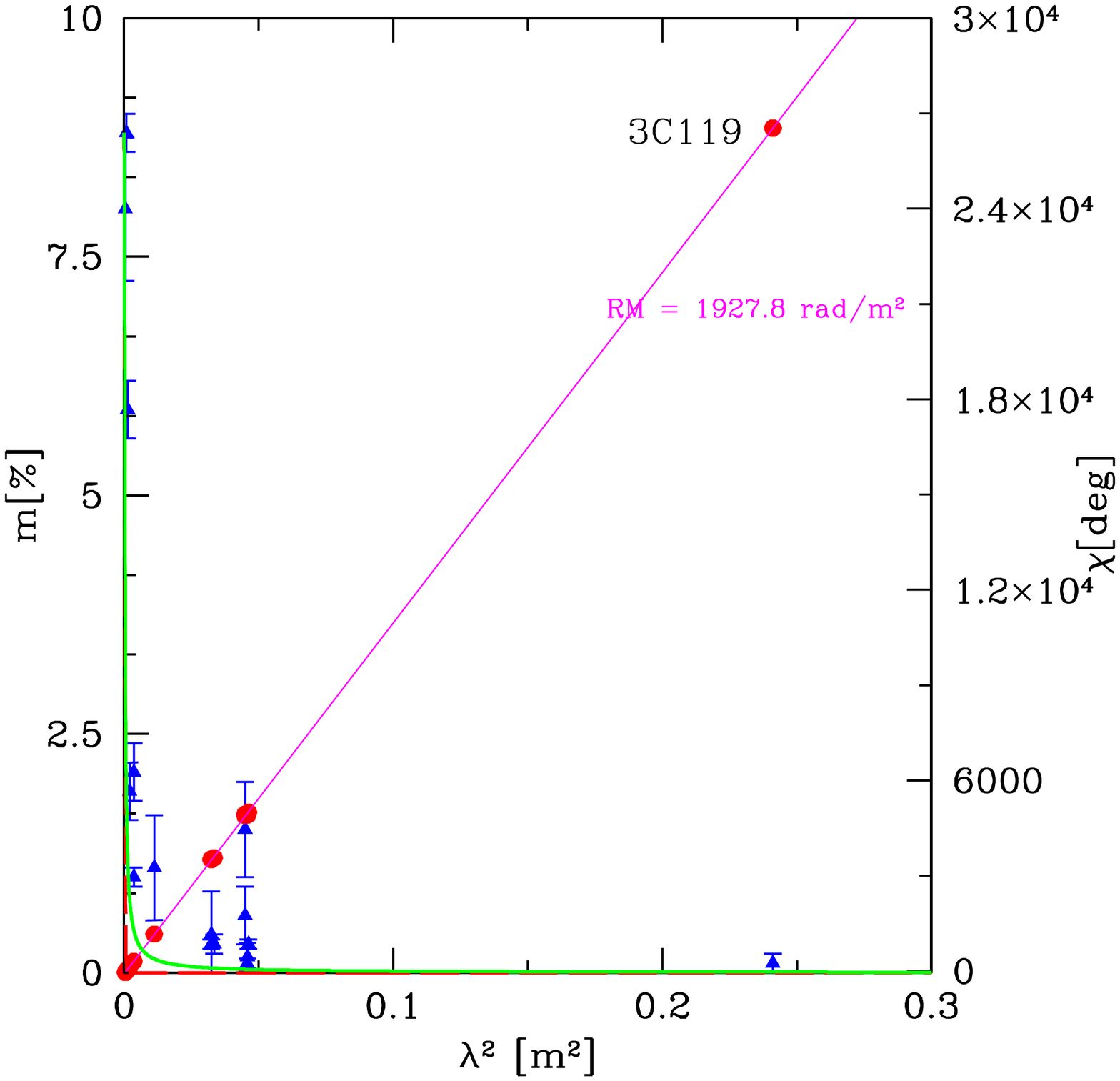}
\includegraphics[width=8cm]{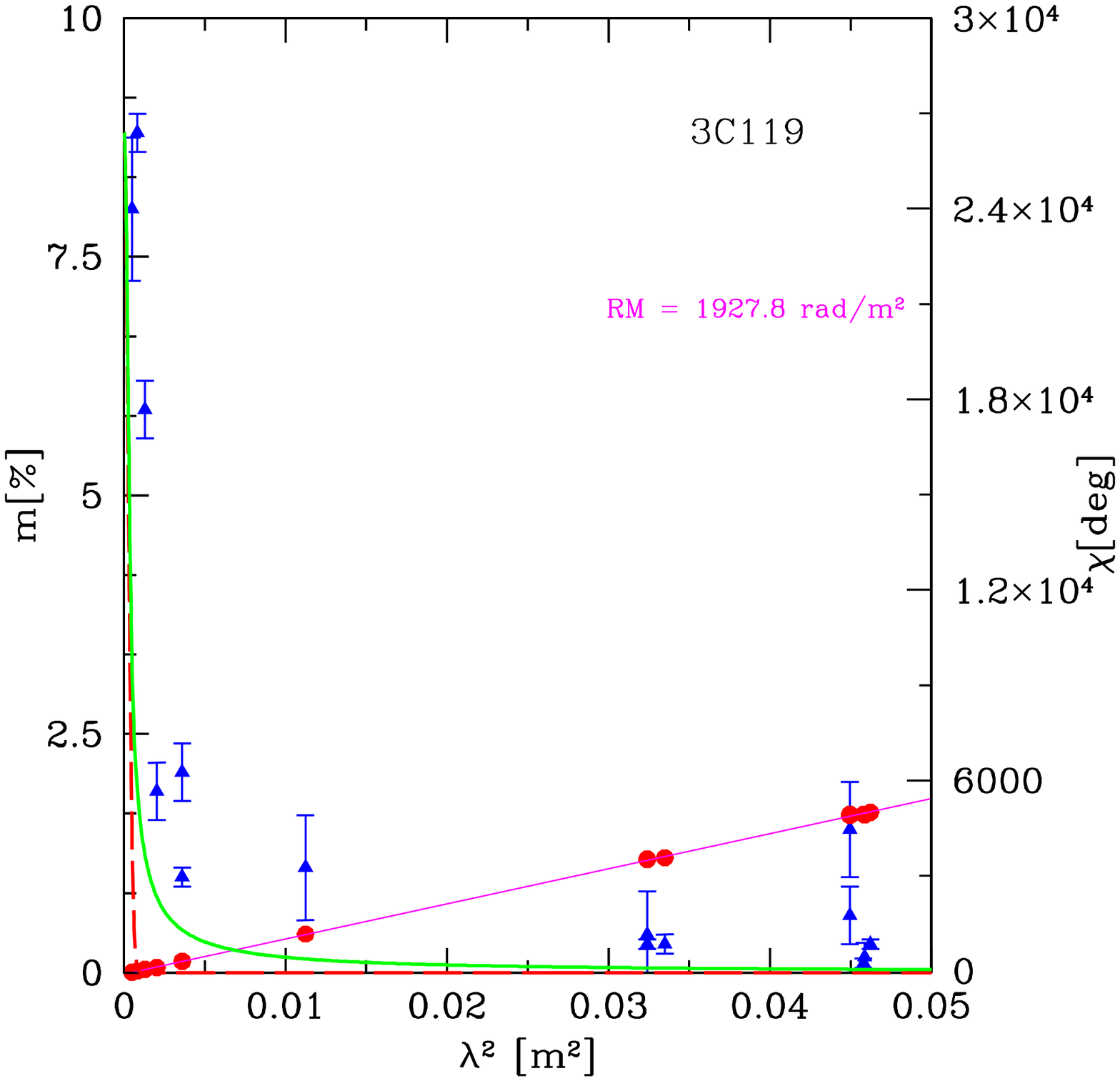}
\caption{Position angles $\chi$ and fractional 
polarisation $m$ for the source 3C119. Layout as in Fig.\,2.}
\end{figure*}
\vspace{-0.3cm}
\begin{figure*}[!hb]
\addtocounter{figure}{+0}
\centering
\includegraphics[width=8cm]{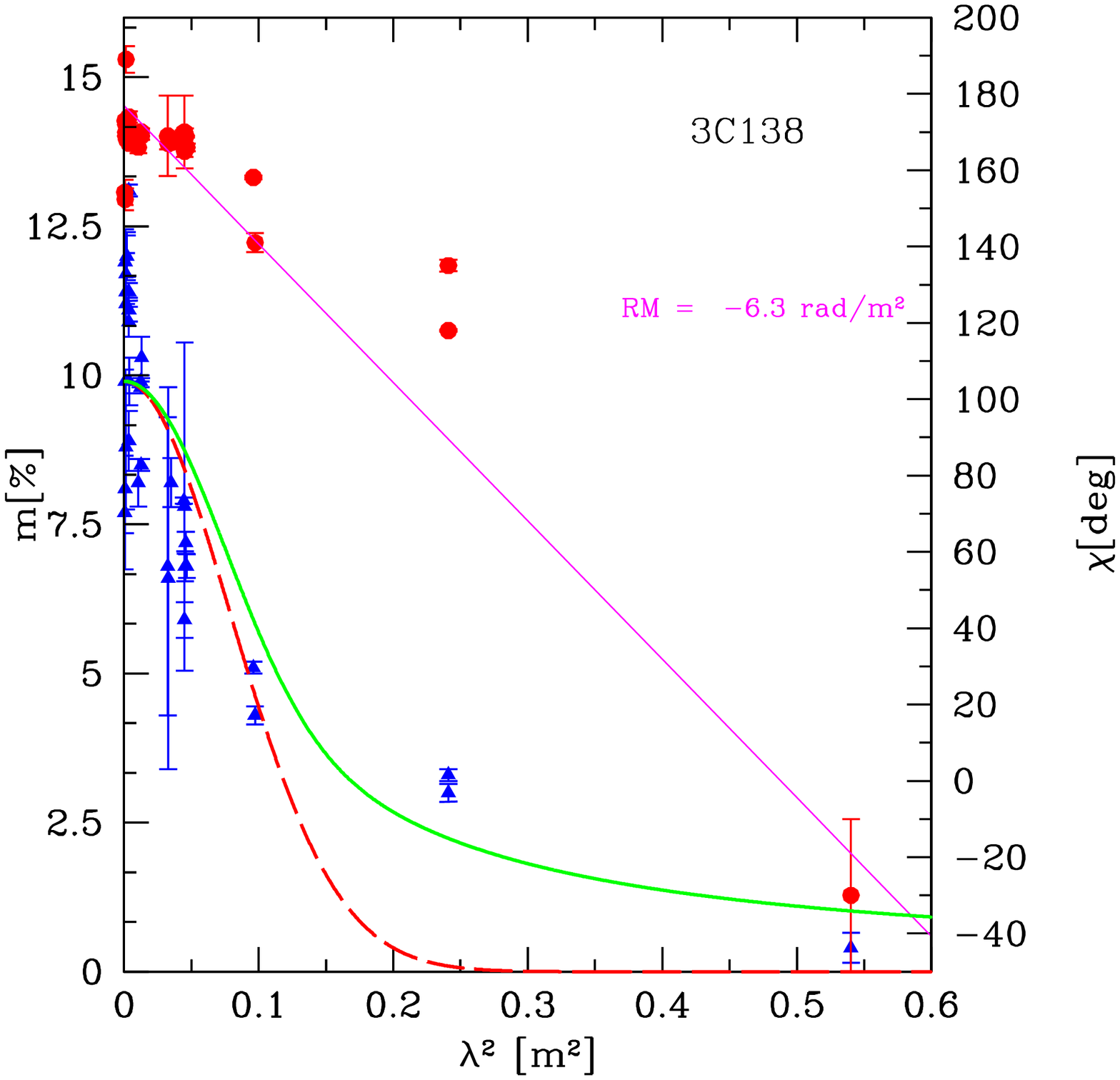}
\includegraphics[width=8cm]{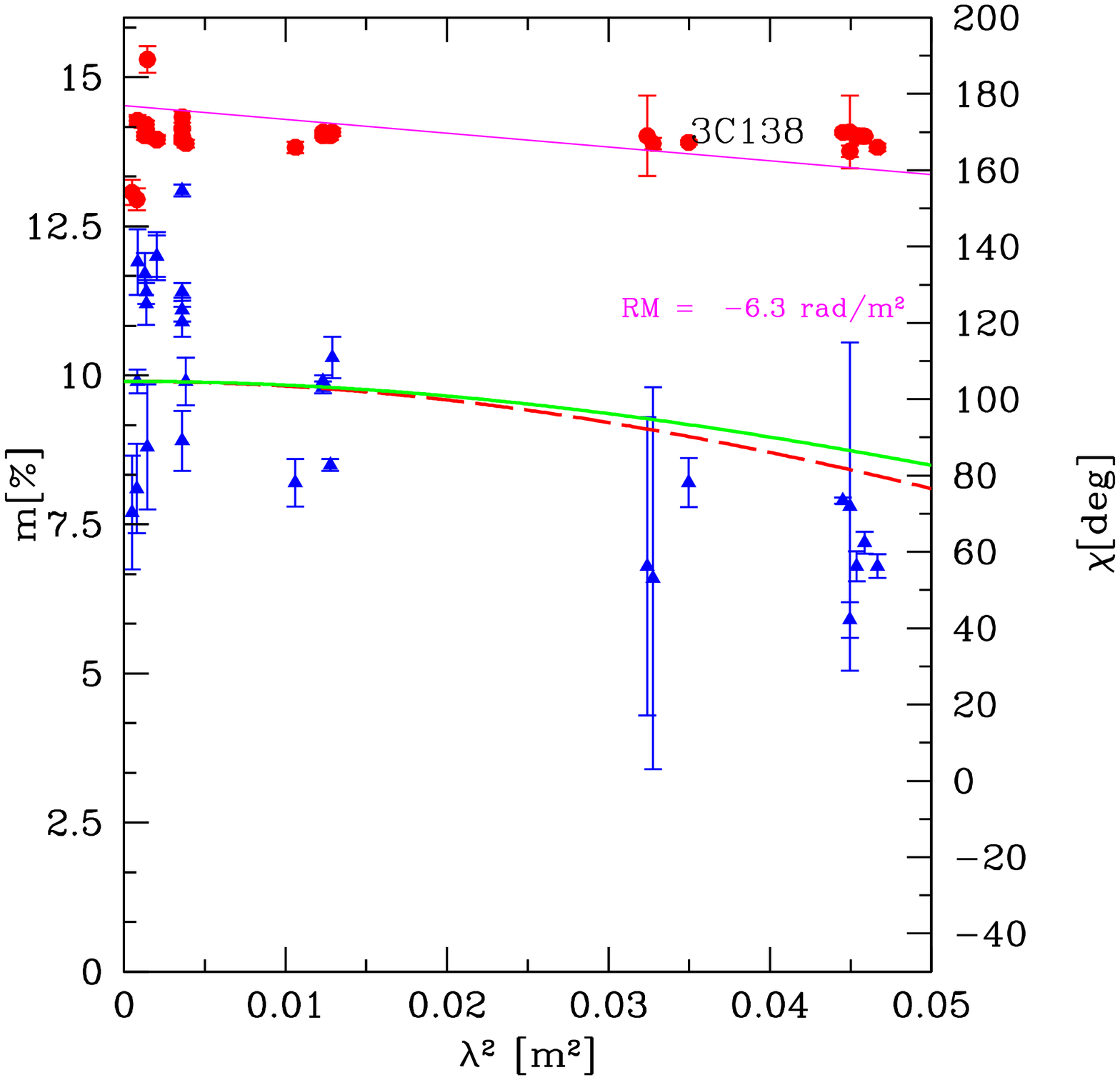}
\caption{Position angles $\chi$ and fractional 
polarisation $m$ for the source 3C138. Layout as in Fig.\,2.}
\end{figure*}
\vspace{-0.3cm}
\begin{figure*}[!hb]
\addtocounter{figure}{+0}
\centering
\includegraphics[width=8cm]{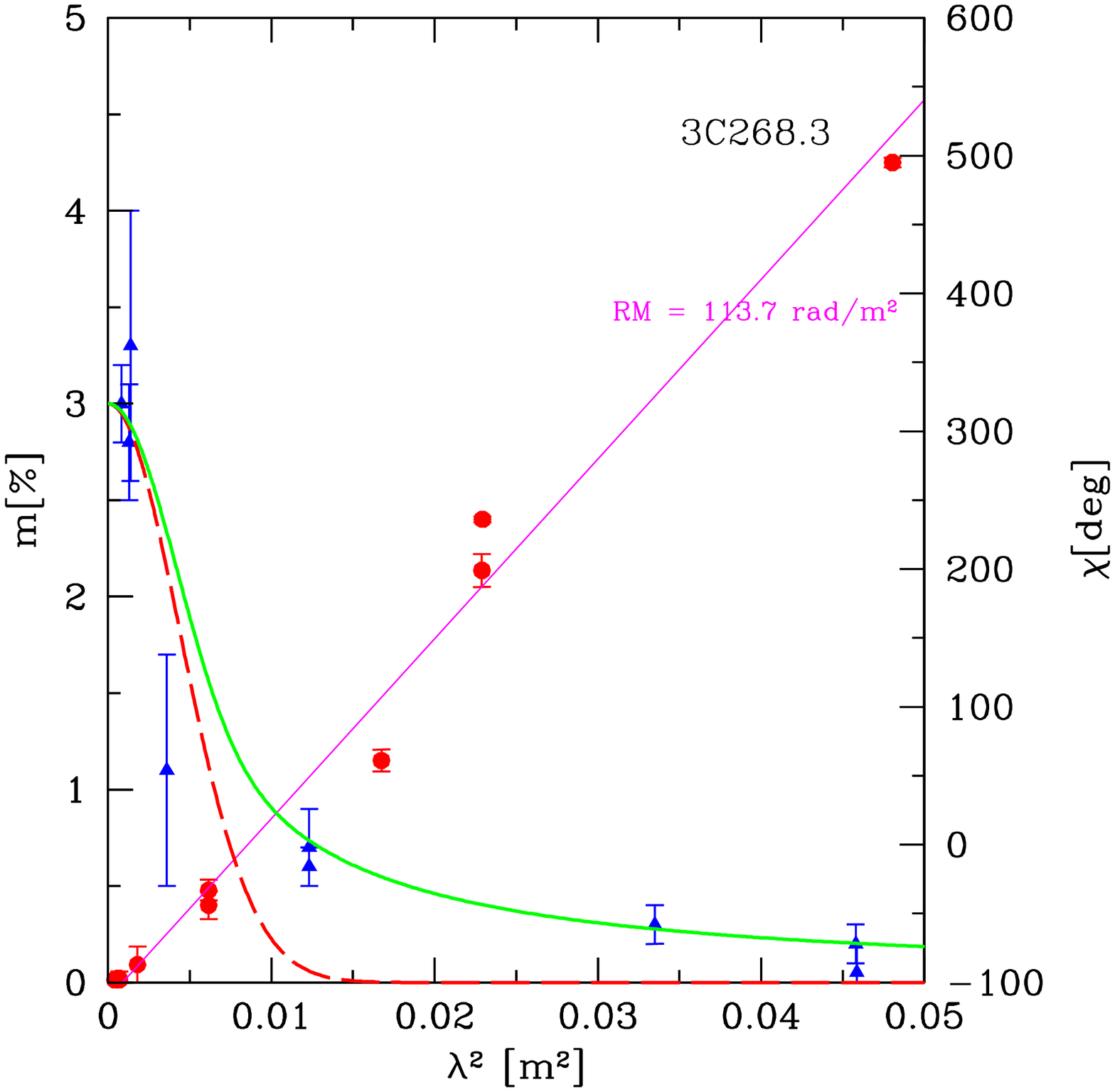}
\caption{Position angles $\chi$ and fractional 
polarisation $m$ for the source 3C268.3. Layout as in Fig.\,2.}
\end{figure*}
\vspace{-0.3cm}
\begin{figure*}[!hb]
\addtocounter{figure}{+0}
\centering
\includegraphics[width=8cm]{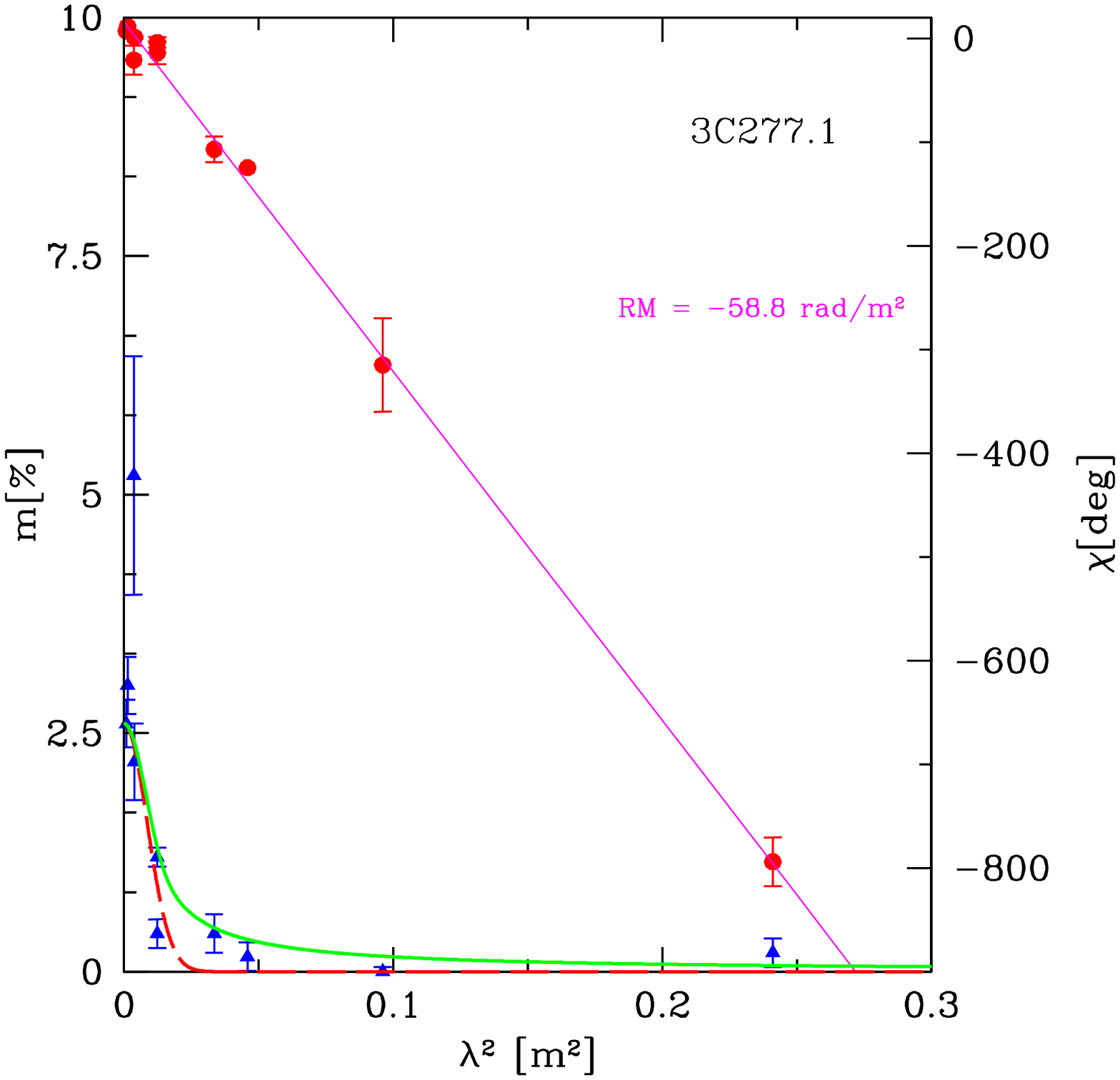}
\includegraphics[width=8cm]{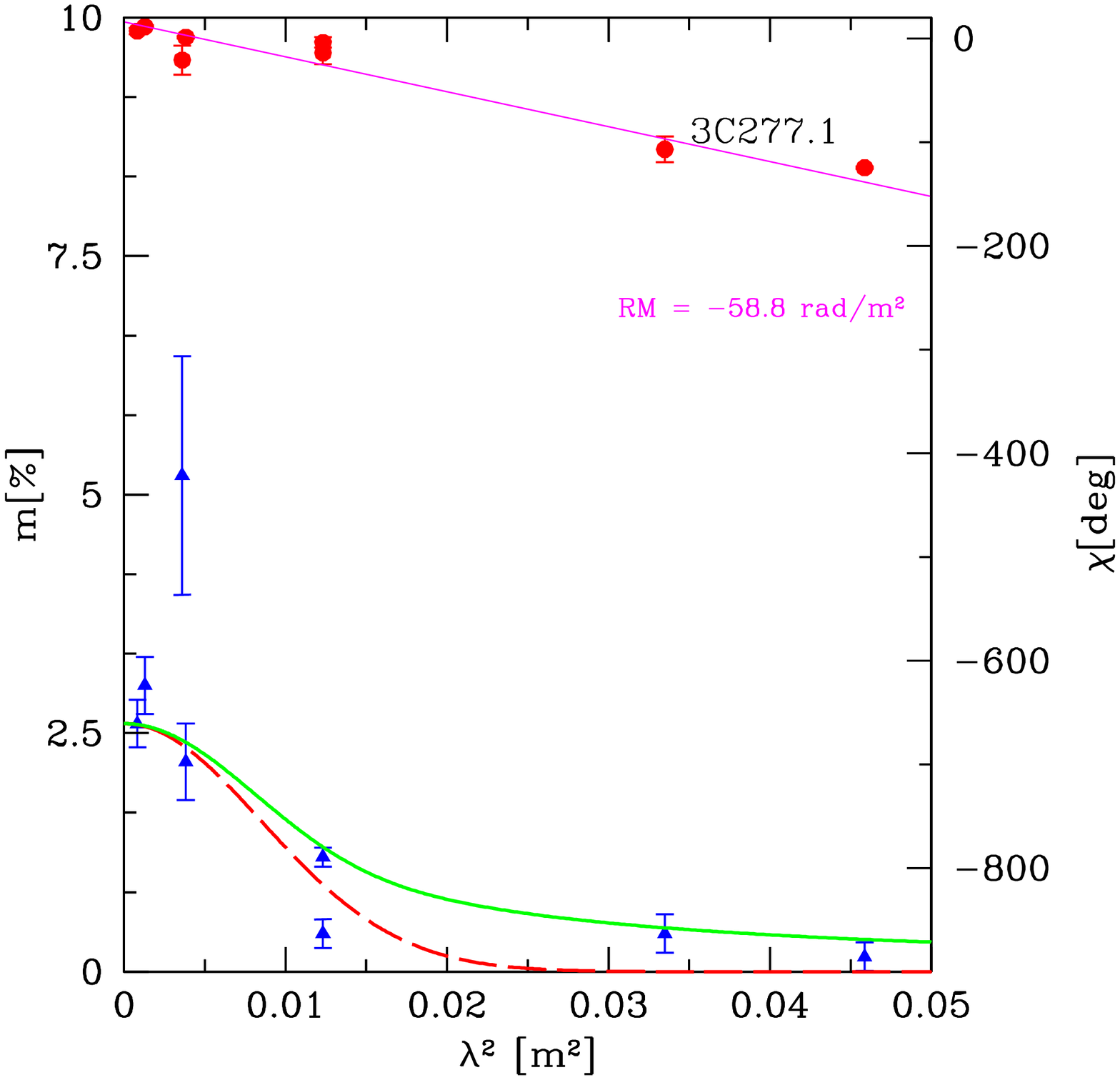}
\caption{Position angles $\chi$ and fractional 
polarisation $m$ for the source 3C277.1. Layout as in Fig.\,2.}
\end{figure*}
\vspace{-0.3cm}
\begin{figure*}[!hb]
\addtocounter{figure}{+0}
\centering
\includegraphics[width=8cm]{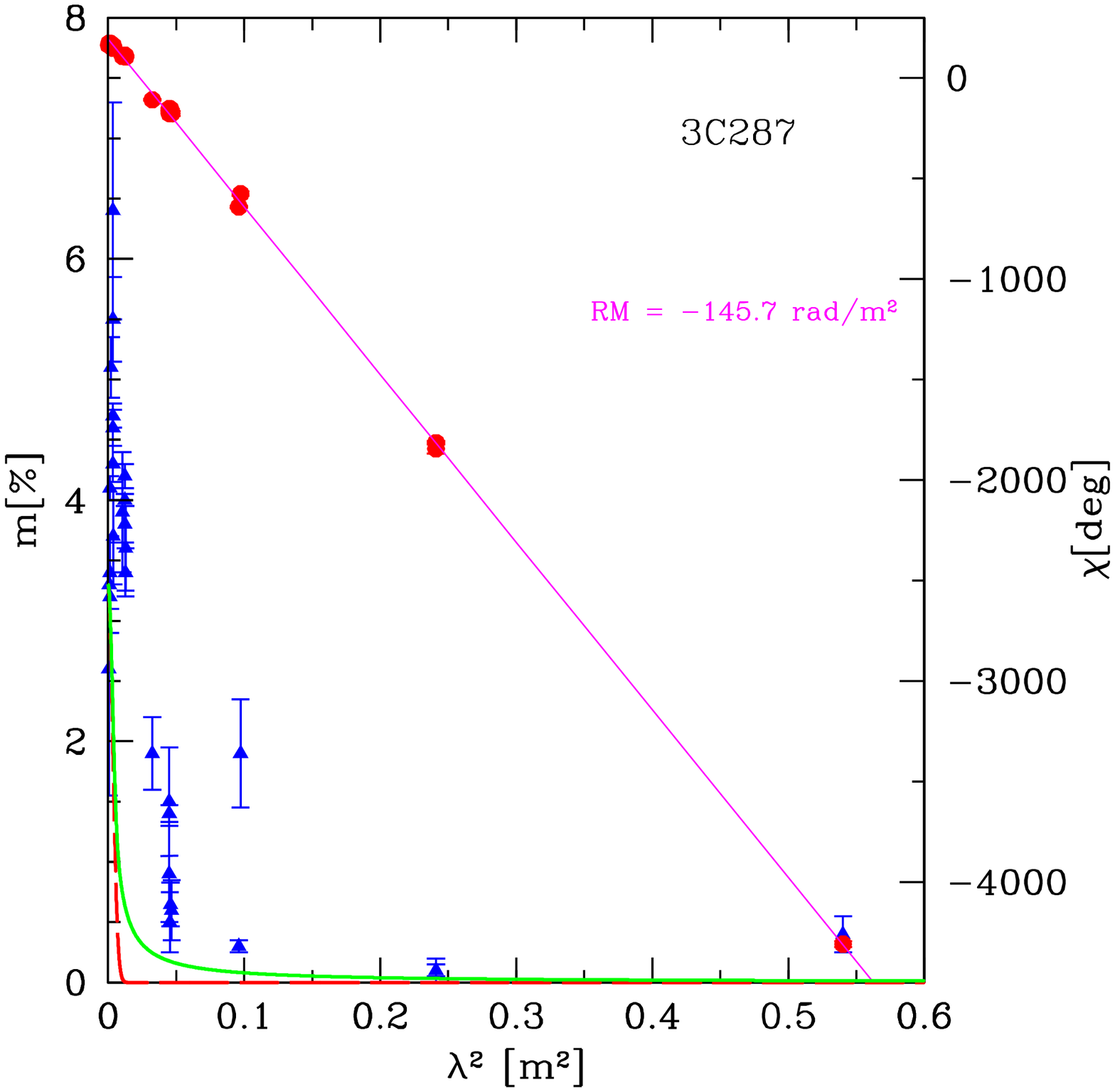}
\includegraphics[width=8cm]{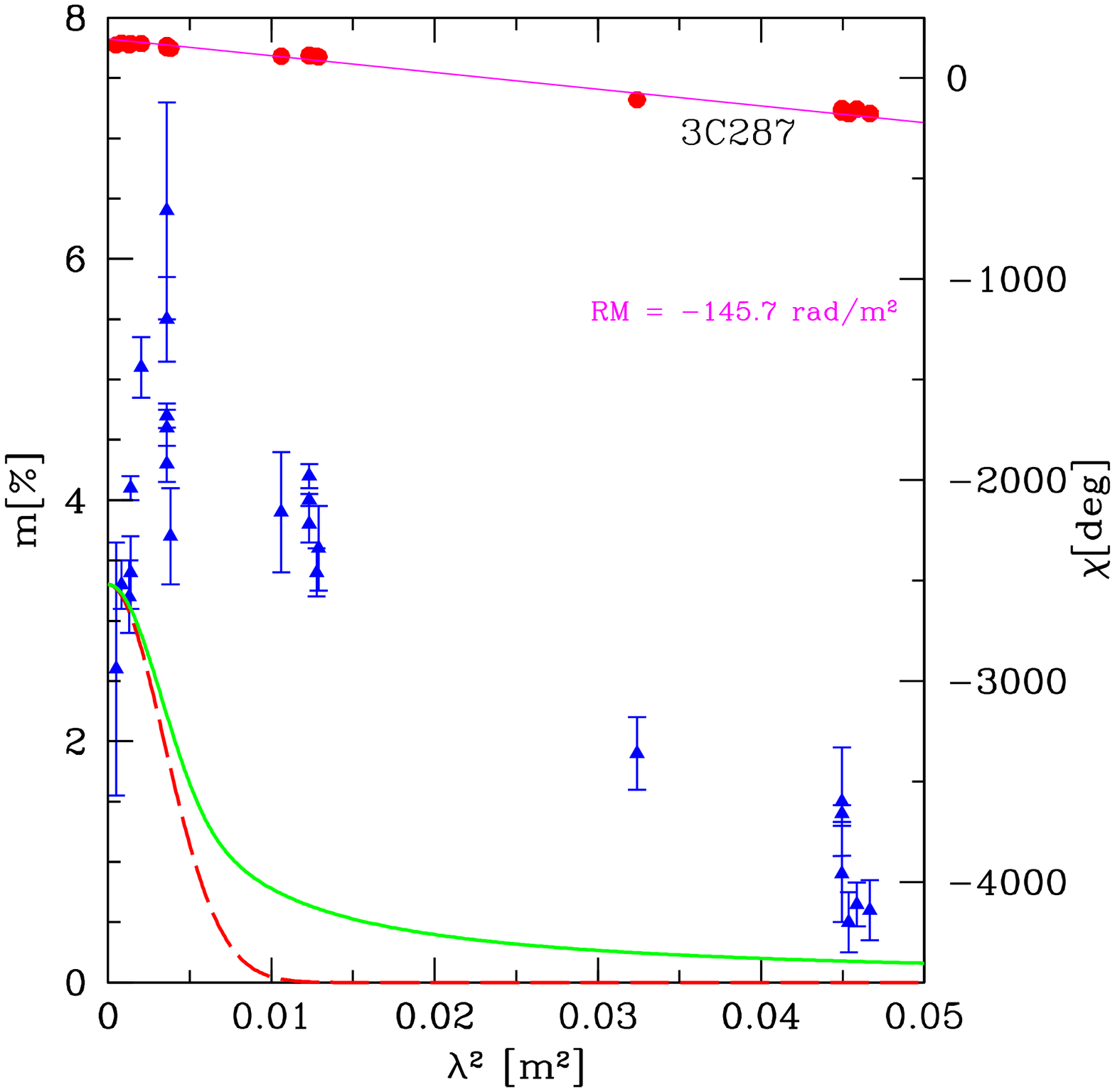}
\caption{Position angles $\chi$ and fractional 
polarisation $m$ for the source 3C277.1. Layout as in Fig.\,2.}

\end{figure*}
\vspace{-0.3cm}
\begin{figure*}[!hb]
\addtocounter{figure}{+0}
\centering
\includegraphics[width=8cm]{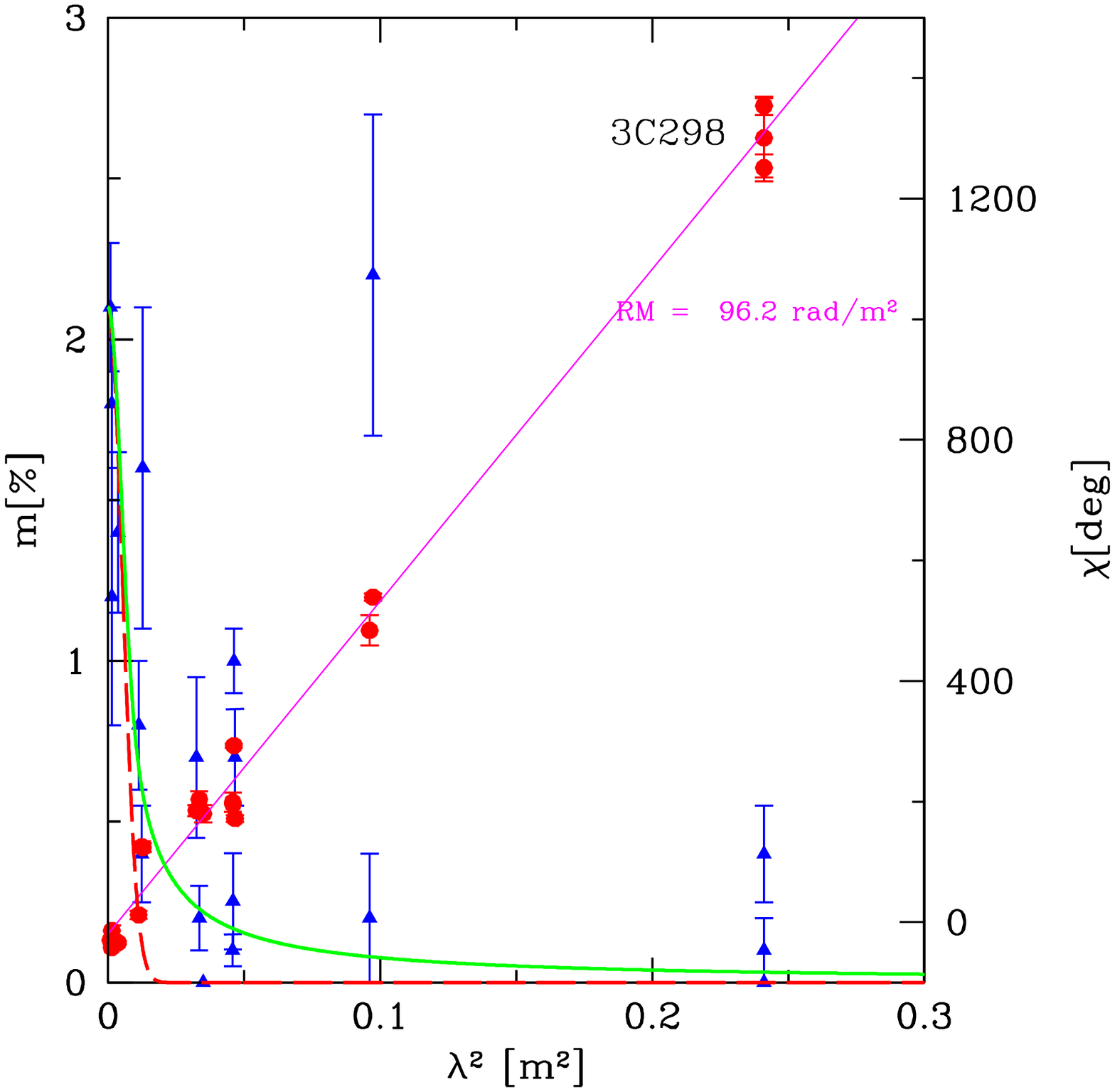}
\includegraphics[width=8cm]{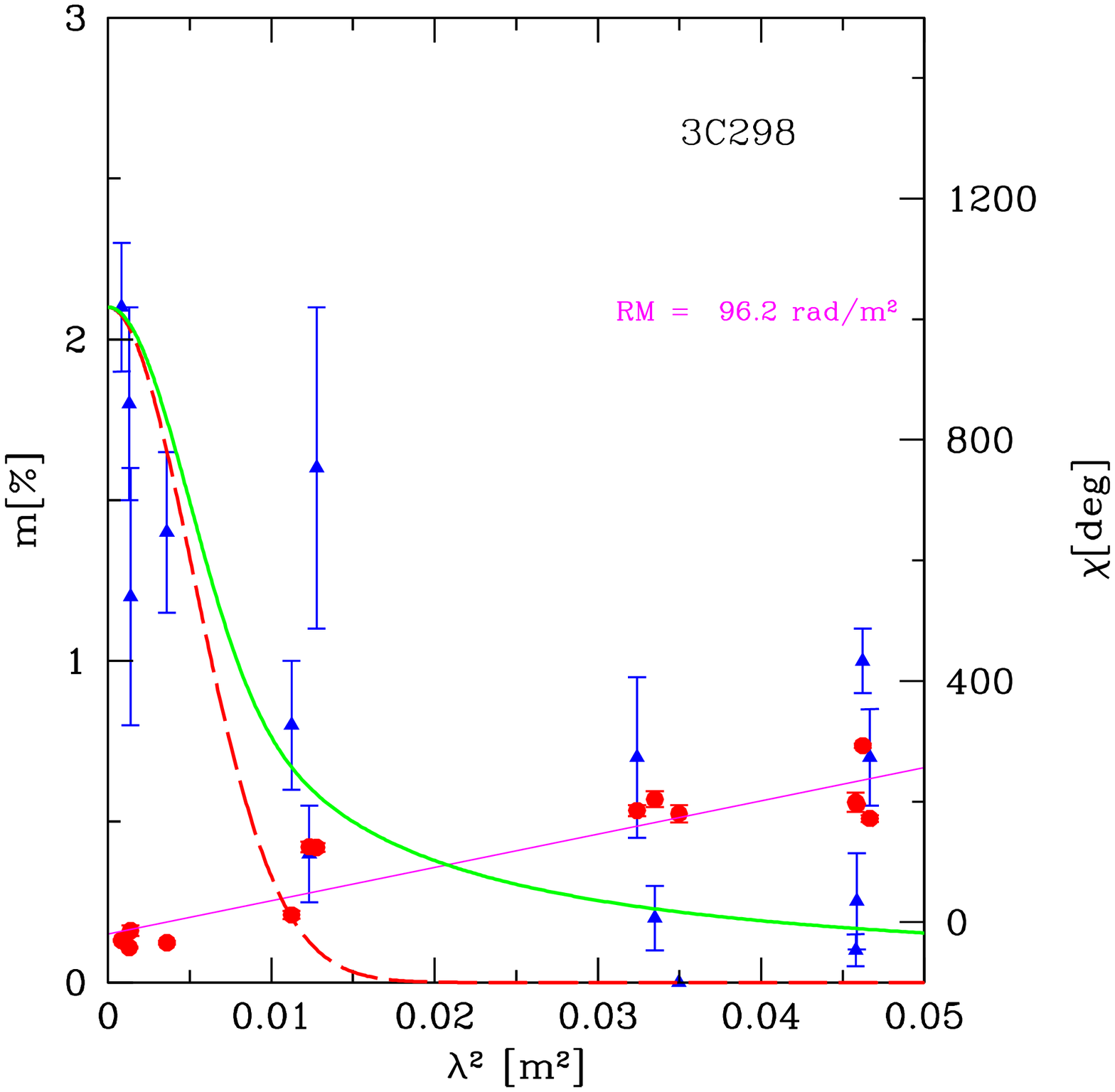}
\caption{Position angles $\chi$ and fractional 
polarisation $m$ for the source 3C277.1. Layout as in Fig.\,2.}
\end{figure*}
\vspace{-0.3cm}
\begin{figure*}[!hb]
\addtocounter{figure}{+0}
\centering
\includegraphics[width=8cm]{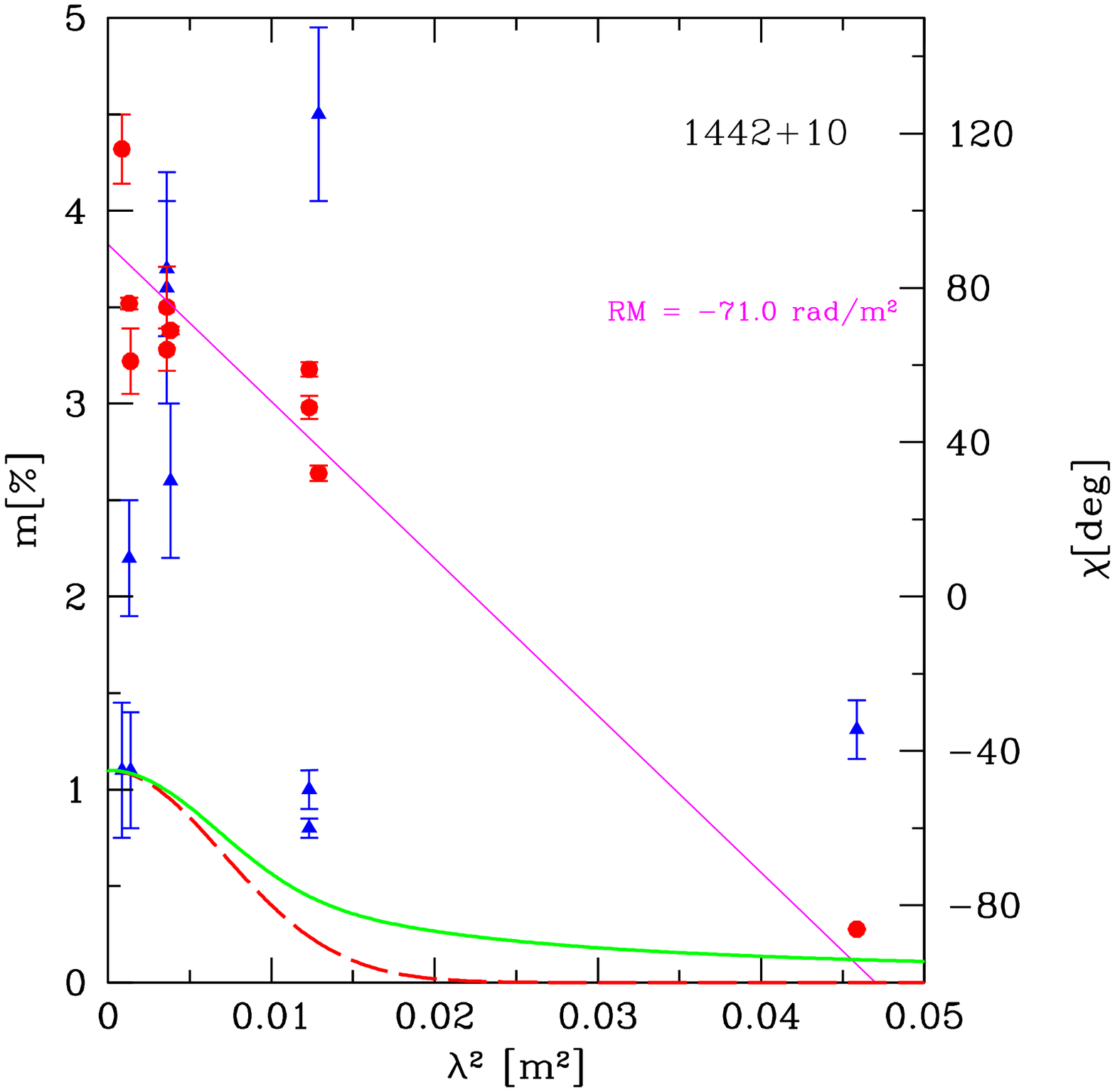}
\caption{Position angles $\chi$ and fractional 
polarisation $m$ for the source 1442+10. Layout as in Fig.\,2.}
\end{figure*}
\vspace{-0.3cm}
\begin{figure*}[!hb]
\addtocounter{figure}{+0}
\centering
\includegraphics[width=8cm]{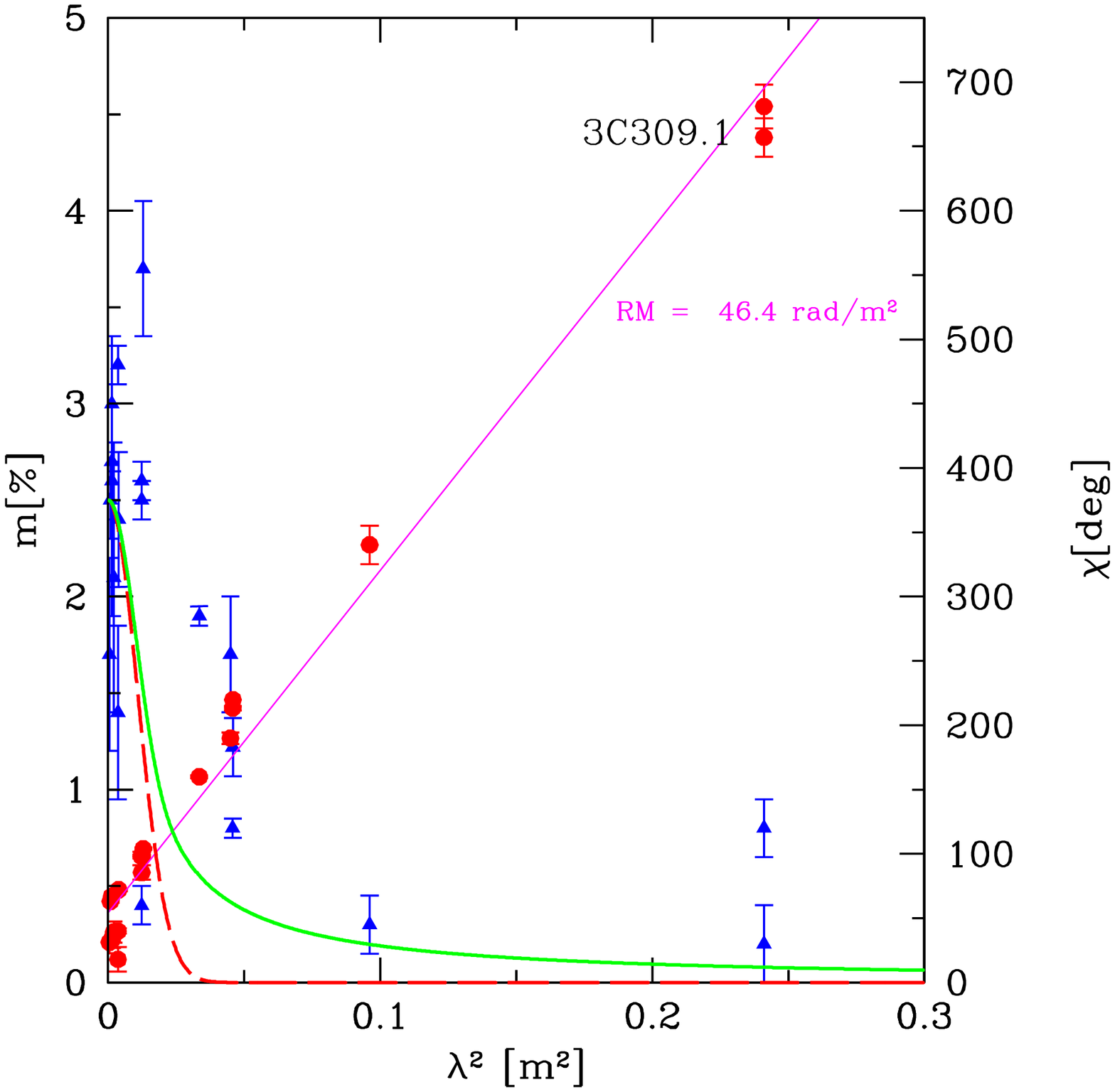}
\includegraphics[width=8cm]{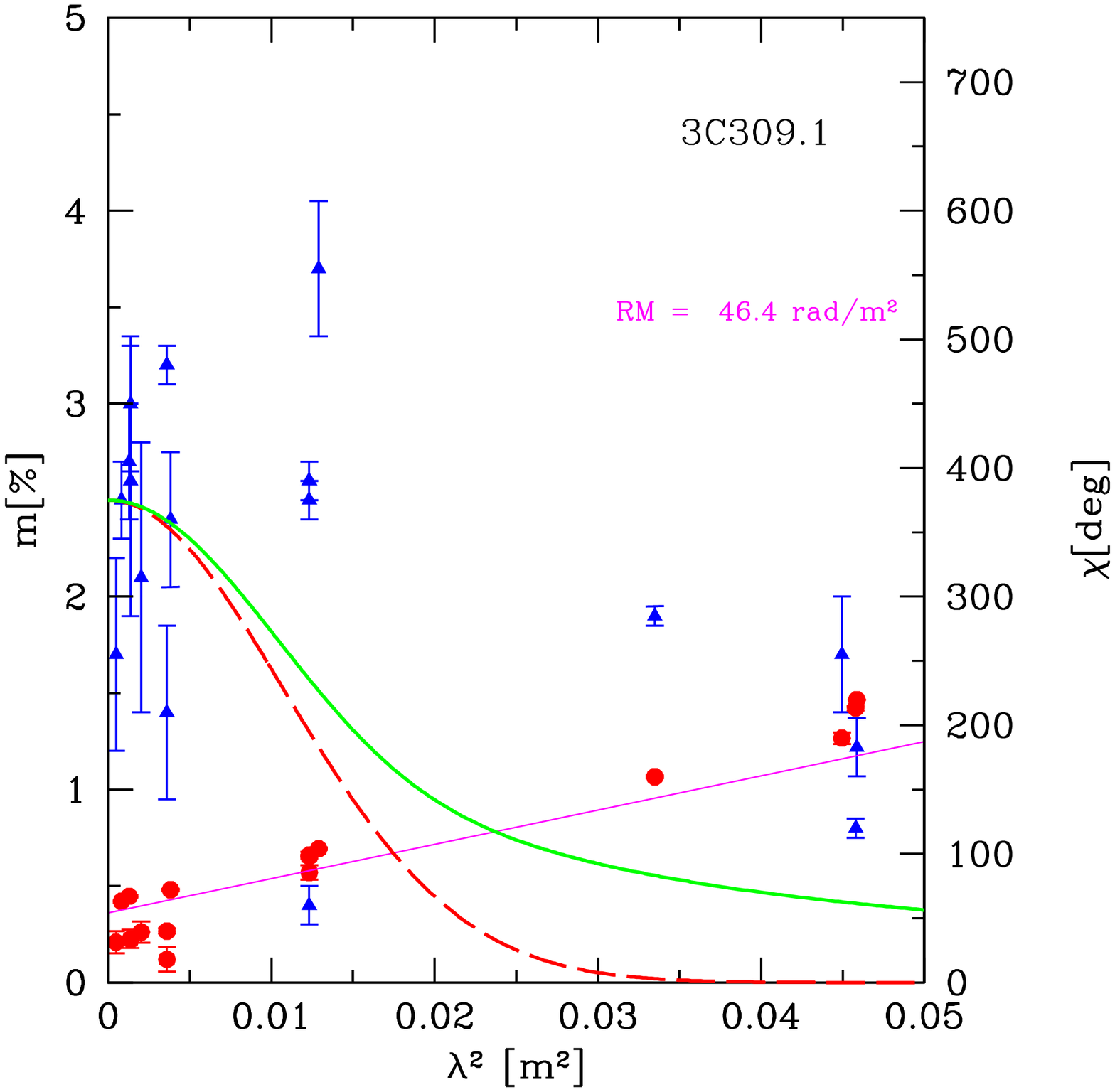}
\caption{Position angles $\chi$ and fractional 
polarisation $m$ for the source 3C309.1. Layout as in Fig.\,2.}
\end{figure*}
\vspace{-0.3cm}
%
%\clearpage
%
\begin{figure*}[!hb]
\addtocounter{figure}{+0}
\centering
\includegraphics[width=8cm]{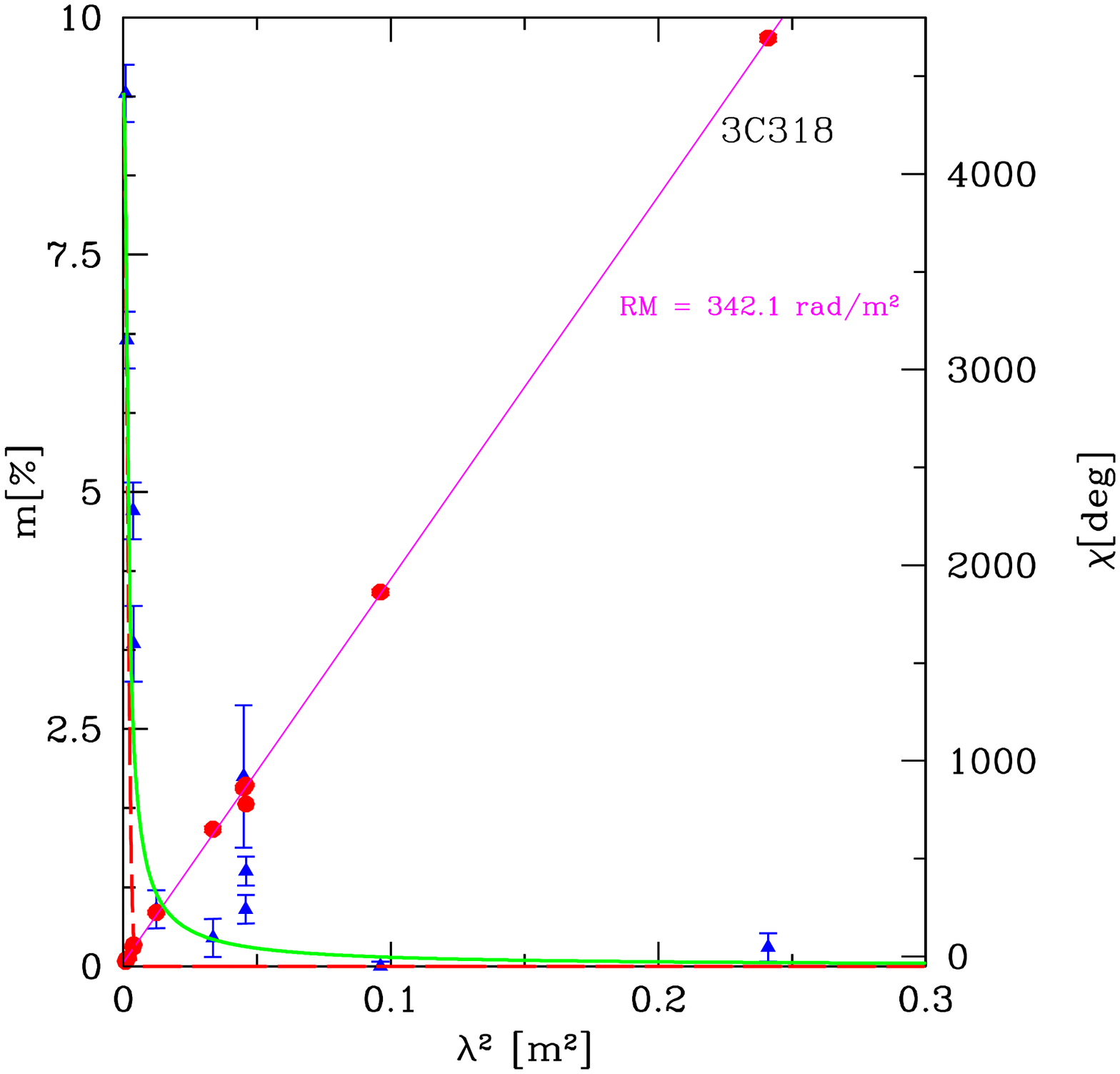}
\includegraphics[width=8cm]{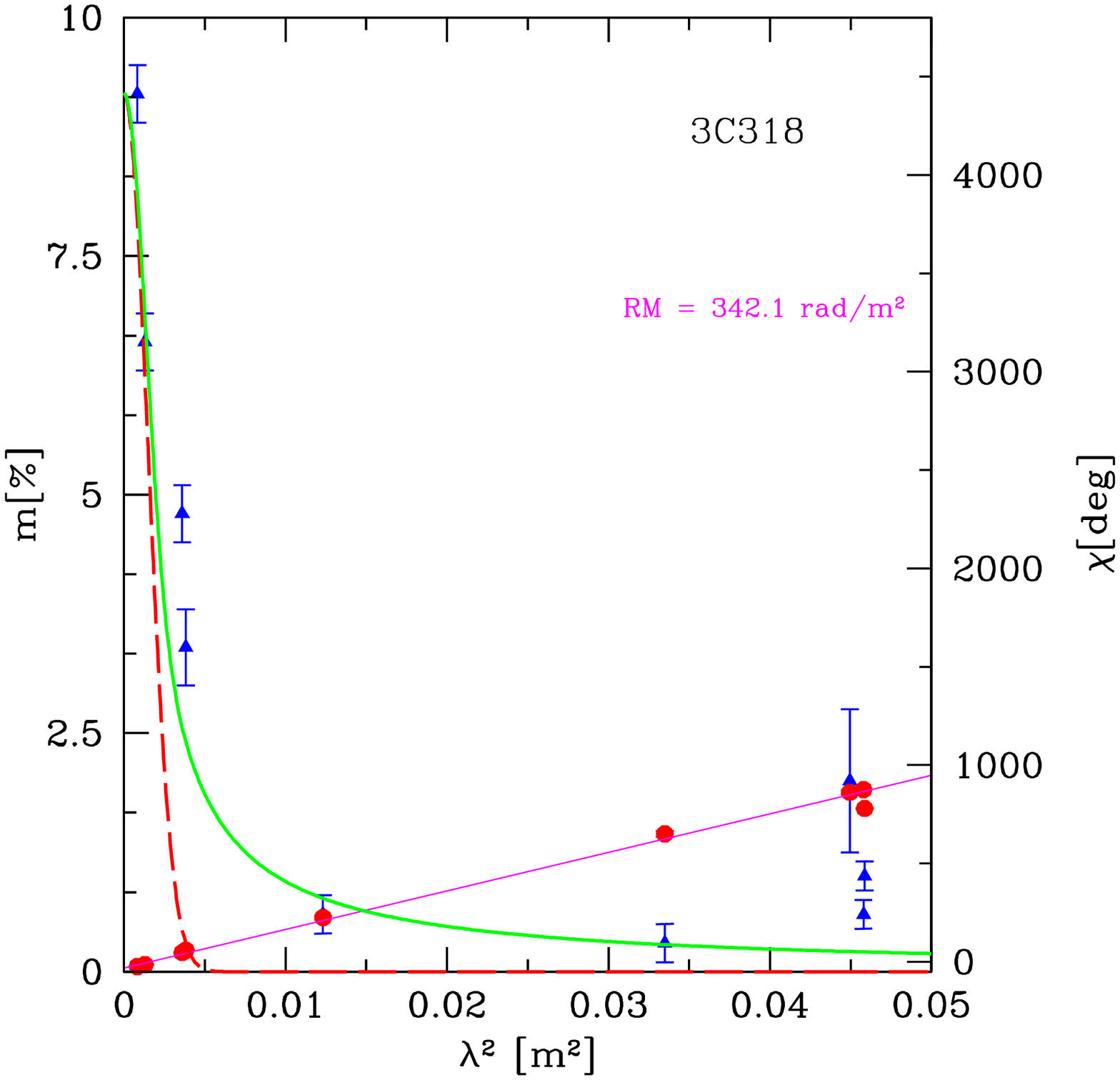}
\caption{Position angles $\chi$ and fractional 
polarisation $m$ for the source 3C318. Layout as in Fig.\,2.}
\end{figure*}
\vspace{-0.3cm}
\begin{figure*}[!hb]
\addtocounter{figure}{+0}
\centering
\includegraphics[width=8cm]{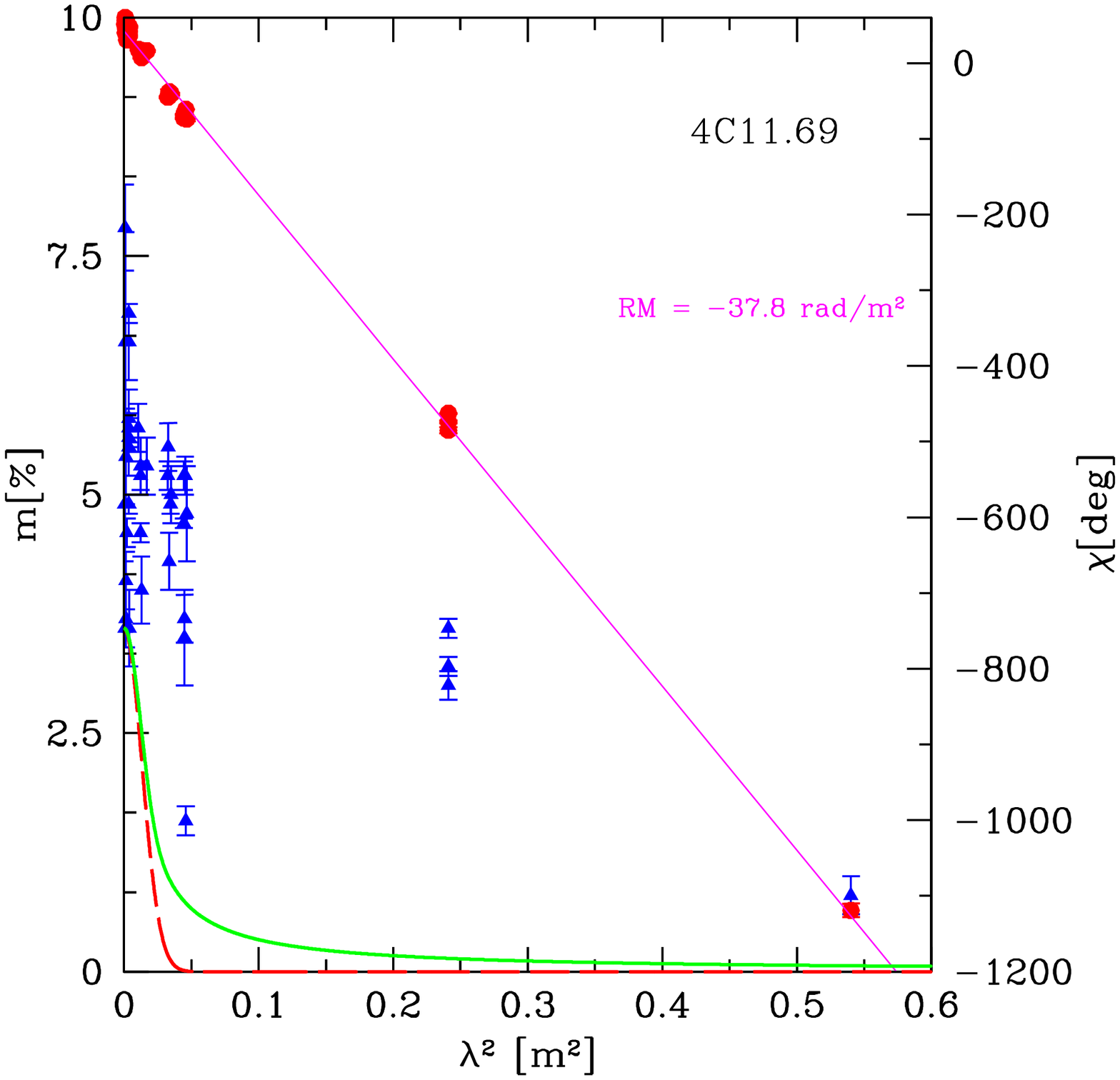}
\includegraphics[width=8cm]{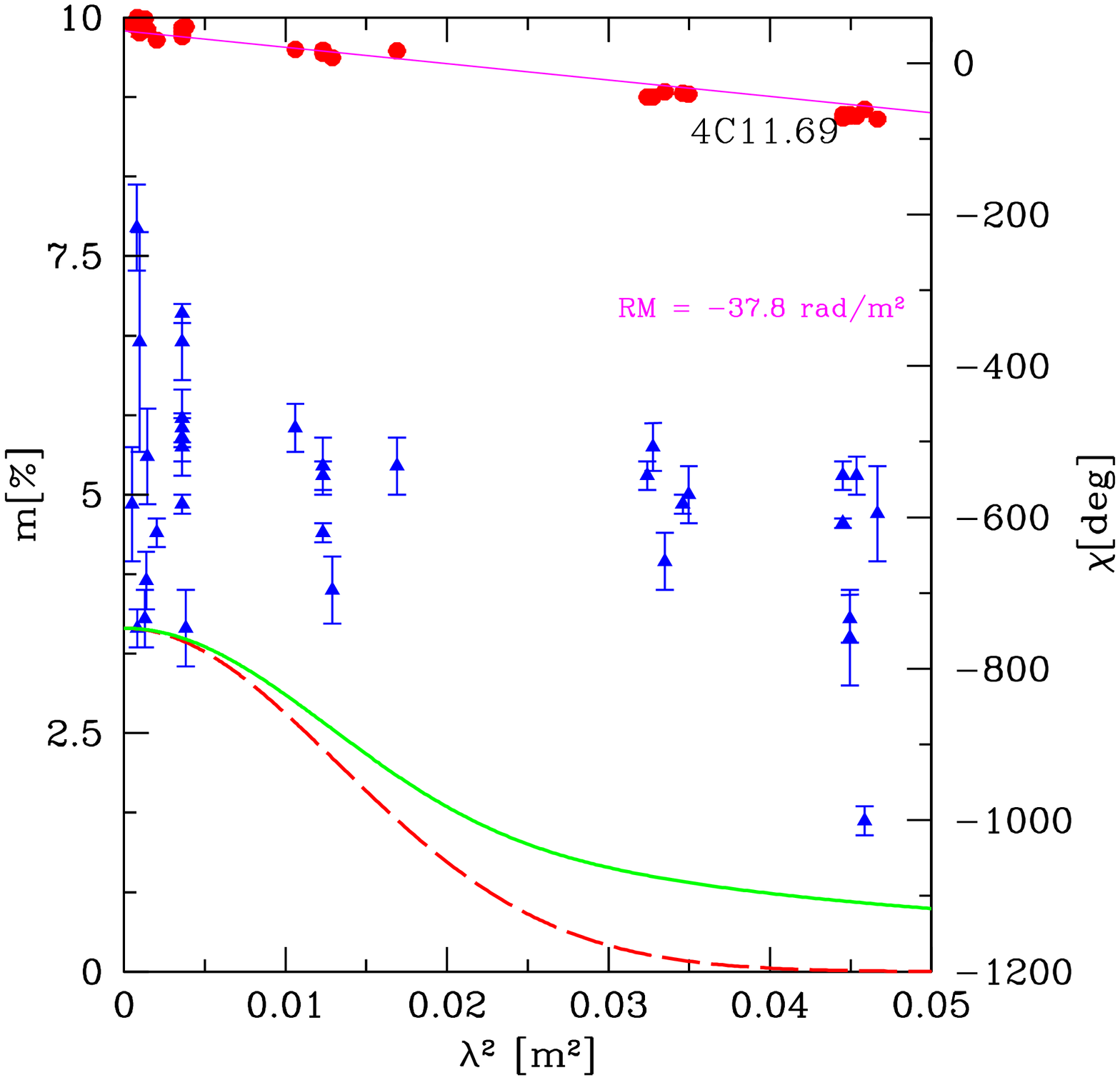}
\caption{Position angles $\chi$ and fractional 
polarisation $m$ for the source 4C11.69. Layout as in Fig.\,2.}
\end{figure*}
\vspace{-0.3cm}
%
%\clearpage
%
\begin{figure*}[!hb]
\addtocounter{figure}{+0}
\centering
\includegraphics[width=8cm]{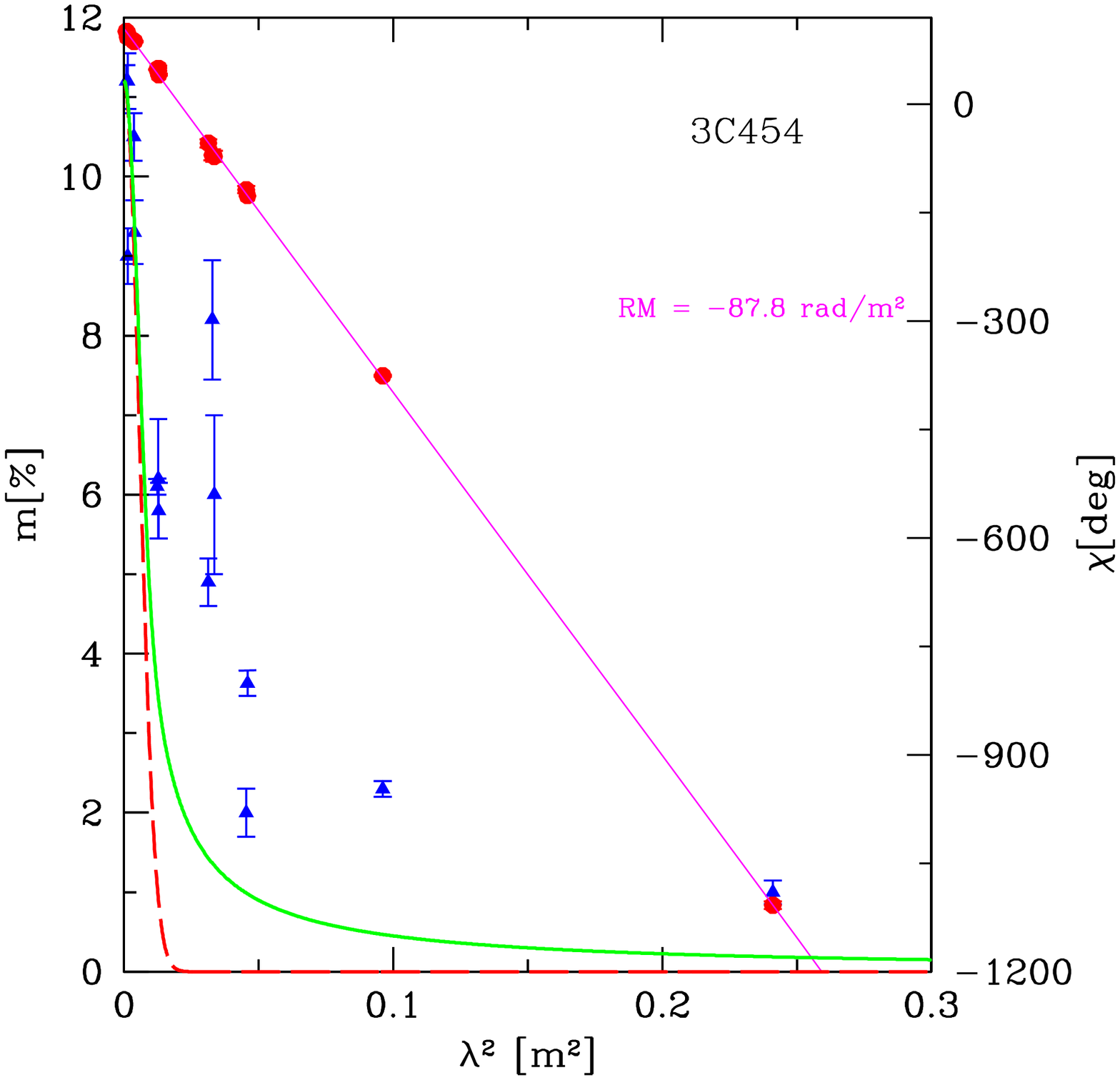}
\includegraphics[width=8cm]{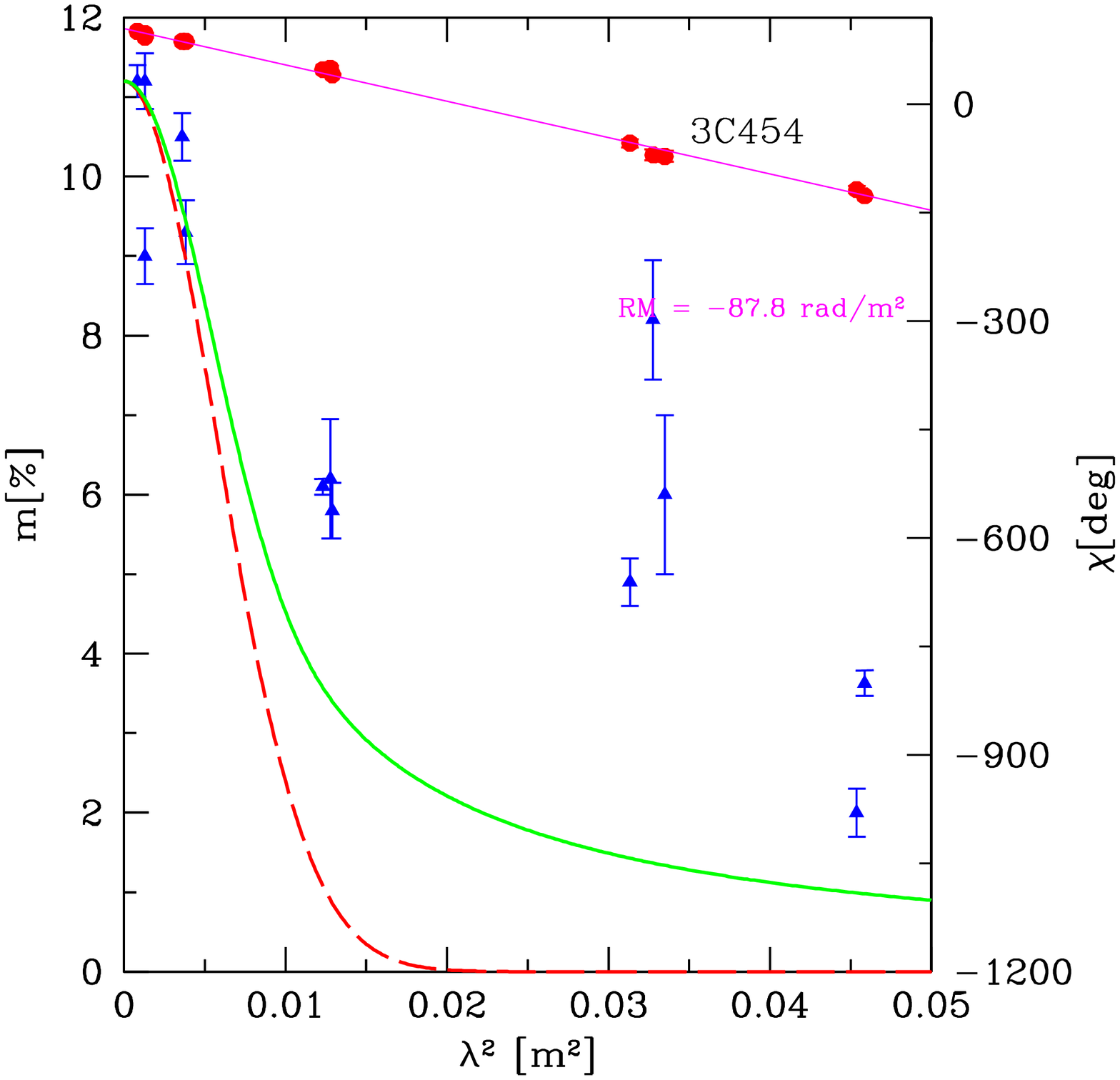}
\caption{Position angles $\chi$ and fractional 
polarisation $m$ for the source 3C454. Layout as in Fig.\,2.}
\end{figure*}
\vspace{-0.3cm}
\clearpage
\begin{figure*}[!hb]
\addtocounter{figure}{+0}
\centering
\includegraphics[width=8cm]{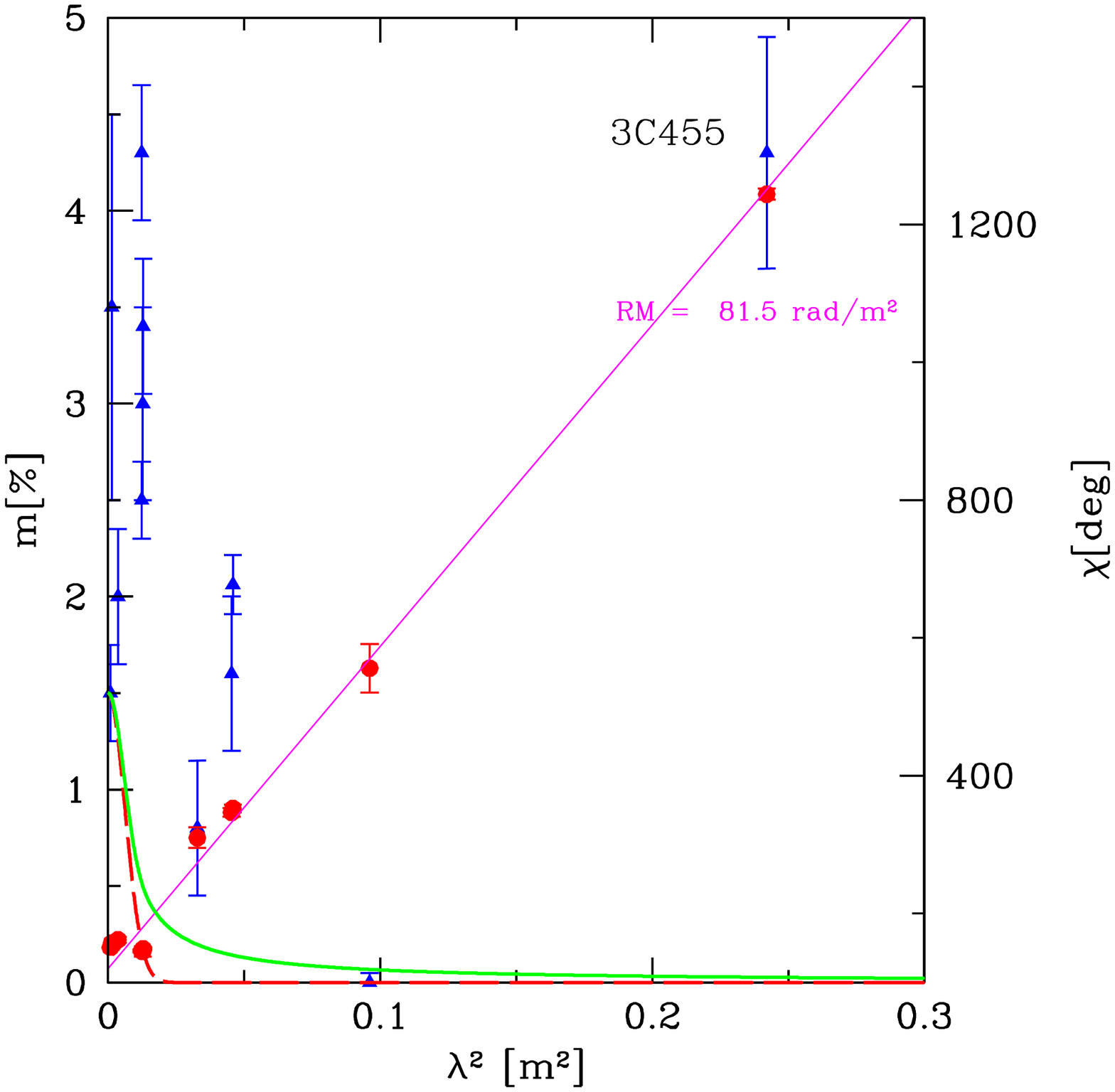}
\includegraphics[width=8cm]{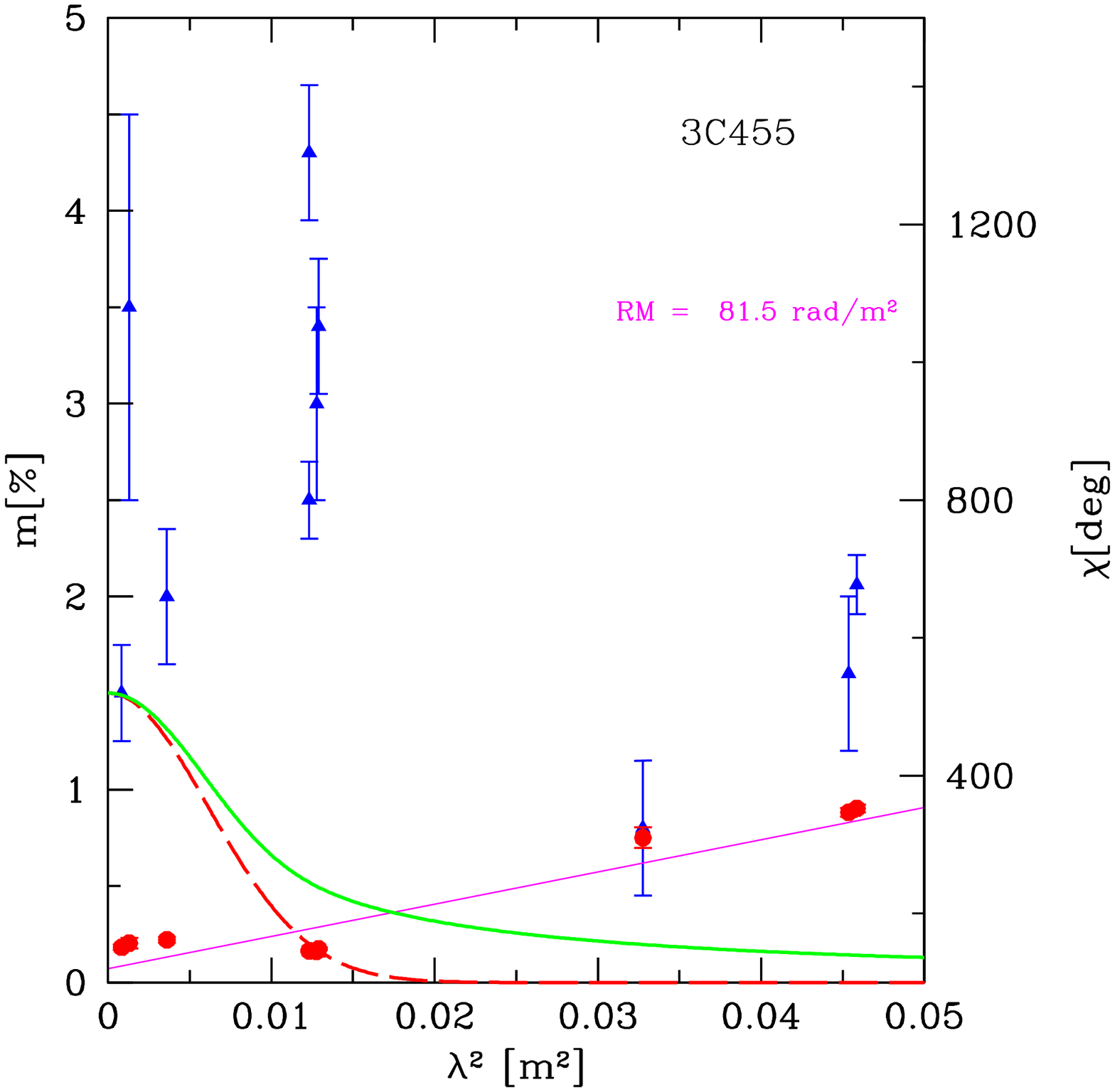}
\caption{Position angles $\chi$ and fractional 
polarisation $m$ for the source 3C455. Layout as in Fig.\,2.}
\end{figure*}
\vspace{-0.3cm}
\clearpage
\newpage

{\bf Appendix 2}
\\
\\
{\bf 2.1 Sources following the Tribble model}

The model proposed by Tribble reproduces the data of about one third of our 
sample, namely of 3C48, 3C67, 3C119, 
3C138, 3C268.3, 3C277.1, and 3C318. 
Sources such as 3C48, 3C67, 3C138, and 3C277.1 also show 
RM$ < \la $ few 10 rad m$^{-2}$, which is an 
indication of an unresolved foreground screen, presumably
the halo of our own Galaxy.

The source 3C119, which has a very high RM and thus
a fast decline in its fractional polarisation, might already be affected by 
significant depolarisation at wavelengths shorter than 2.8\,cm. A higher m$_0$
would lead to depolarisation according to the Tribble law.
\\
\\
{\bf 2.2 Sources with an indication of repolarisation}

Eight sources (3C43, 0319+12, 3C287, 1442+10,
3C309.1, 4C11.69, 3C454, and 3C455) show indications of repolarisation,
i.e., an increase in fractional polarisation with decreasing frequency,
generally at short wavelengths $\la$10\,cm; this corresponds to a relatively 
strong
increase in fractional polarisation followed by a decline that can still
be described by the Tribble law. This effect is visible despite possible 
instrumental effects of different telescopes and also considering that 
observations were made at different epochs.
In three of these sources (3C309.1, 4C11.69, 3C454), the repolarisation 
effect is indeed confused with possible time variability. Focusing on
our simultaneous measurements only, we still find constant or slightly
increasing fractional polarisation with increasing wavelength, which is
not predicted by any depolarisation model. 
In particular, the galaxy 3C455 shows a measured repolarisation 
greater than the
3$\sigma$ level between 10.45\,GHz and 2.64\,GHz together with
a small value of the RM$_{rf}$ (194 rad m$^{-2}$), which is an indication of 
the influence of a foreground screen, possibly an extended cloud with 
[OII] emission detected by Hes et al. (1996). 
A plausible mechanism
would be the effect of shear layers caused by the interaction between the 
surface of an expanding source and the surrounding medium (Burn 1966).

The source 3C455 was observed with the VLA at 8.35\,GHz by Bogers et al. 
(1994).
It shows a triple structure with the indication of a jet joining the
core with the south western lobe. The three components are almost aligned along
the source major axis extending up to about 4$\arcsec$. However, 3C455 
appears slightly resolved by the NVSS, which has a restoring beam of 
45$\arcsec$,
suggesting that the three components imaged by Bogers et al. are 
actually embedded in a more extended region of low brightness emission. 
This can be seen in the image available in the
VLA Low-Frequency Sky Survey (VLSS; Cohen et al. 2007) at 74\,MHz, which shows 
an even more extended structure of about 2$\arcmin$ in size. The existence of 
this extended emission is supported by the source spectral index, 
which indicates an upturn towards higher flux density above $\sim$100\,MHz. 
Therefore, a second 
possible interpretation is that by observing at 2.64\,GHz or lower 
frequencies, we have integrated the 
flux density and polarised flux density from that region. 
At 10.45\,GHz, the steep spectrum extended structure is below the detection
limit.

A similar case of repolarisation at a lower frequency was pointed out by 
Montenegro-Montes et al. (2008) for the source 1159+01.
\\
\\
{\bf 2.3 Polarisation variability}

Finally, 3C298 could exhibit time variability in its fractional
polarisation. Unfortunately, the current data base does not provide a 
sufficient
number of simultaneous measurements to prove this effect, which was reported 
for example by Aller et al. (2003) for the source 3C147.
\end{document}